 \documentclass[preprint,3p,times,twocolumn]{elsarticle}

\usepackage{graphicx}      
\usepackage{natbib}        
\usepackage{amsmath}
\usepackage{amsfonts}
\usepackage{amssymb}
\usepackage{upgreek}
\usepackage{nicefrac}
\usepackage{mathrsfs}
\usepackage[breaklinks=true,colorlinks=true,allcolors=cyan]{hyperref}
\usepackage{xurl}
\newcommand{\Sabs}[1]{\left\lceil #1 \right\rfloor}

\newcommand{\norm}[1]{\left|\left| #1 \right|\right|}

\newcommand{\sign}{\textup{sign}}

\newtheorem{thm}{Theorem}
\newtheorem{lem}{Lemma}
\newtheorem{dfn}{Definition}
\newdefinition{rem}{Remark}
\newproof{pf}{Proof}
\newtheorem{assumption}{Assumption}

\journal{Journal}

\begin{document}

\begin{frontmatter}



\title{Insights of using Control Theory for minimizing Induced Seismicity in Underground Reservoirs}

\author[Gem]{Diego Guti\'errez-Oribio}
\ead{diego.gutierrez-oribio@ec-nantes.fr}
\author[Gem]{Ioannis Stefanou\corref{cor1}}
\ead{ioannis.stefanou@ec-nantes.fr}

\cortext[cor1]{Corresponding author.}

\affiliation[Gem]{organization={Nantes Universite, Ecole Centrale Nantes, CNRS, GeM},
             addressline={UMR 6183, F-44000},
             city={Nantes},
             country={France}}
                          
\begin{abstract}

Deep Geothermal Energy, Carbon Capture, and Storage and Hydrogen Storage have significant potential to meet the large-scale needs of the energy sector and reduce the CO$_2$ emissions. However, the injection of fluids into the earth's crust, upon which these activities rely, can lead to the formation of new seismogenic faults or the reactivation of existing ones, thereby causing earthquakes. In this study, we propose a novel approach based on control theory to address this issue. First, we obtain a simplified model of induced seismicity due to fluid injections in an underground reservoir using a diffusion equation in three dimensions. Then, we design a robust tracking control approach to force the seismicity rate to follow desired references. In this way, the induced seismicity is minimized while ensuring fluid circulation for the needs of renewable energy production and storage. The designed control guarantees the achievement of the control objectives even in the presence of system uncertainties and unknown dynamics. Finally, we present simulations of a simplified geothermal reservoir under different scenarios of energy demand to show the reliability and performance of the control approach, opening new perspectives for field experiments based on real-time regulators.

\end{abstract}



\begin{keyword}

Energy geotechnics \sep Geothermal energy \sep Energy and energy product storage \sep Earthquake prevention \sep Induced seismicity \sep Robust control 

\end{keyword}

\end{frontmatter}


\section{INTRODUCTION}

Deep Geothermal Energy, Carbon Capture and Storage, and Hydrogen Storage show promising potential in addressing the substantial energy sector demands while mitigating CO$_2$ emissions. However, their effectiveness relies on injecting fluids into the Earth's crust, a process that may potentially induce earthquakes \cite{b:10.1785/0220170112, b:Rubinstein-Mahani-2015, b:10.1002/2016RG000542}. The occurrence of induced seismicity poses a significant threat to the feasibility of projects employing these techniques. This concern has led to the closure of several geothermal plants worldwide, such as those in Alsace, France, in 2020 \cite{b:Maheux_FBleu, b:Stey_LeMonde}, Pohang, South Korea, in 2019 \cite{b:Sun_Hank, b:Zastrow-2019}, and Basel, Switzerland, in 2009 \cite{b:10.1785/gssrl.80.5.784, b:Glanz_NYT}.

Earthquakes initiate when there is a sudden release of significant elastic energy stored within the Earth's crust due to abrupt sliding along fault lines \cite{b:Scholz-2002, b:Kanamori-Brodsky-2004}. The injection of fluids can lead to the formation of new seismogenic faults or the reactivation of existing ones, which cause earthquakes \cite{b:Rubinstein-Mahani-2015, b:10.1002/2016RG000542, b:Keranen-Savage-Abers-Cochran-2013}. More specifically, elevated fluid pressures at depth amplify the amount of elastic energy accumulated in the Earth's crust while reducing the friction along faults. As a result, the likelihood of earthquakes significantly increases, even in regions that are typically considered to have low seismic potential (see \cite{b:Rubinstein-Mahani-2015}, \cite{b:Zastrow-2019}, and \cite{b:Keranen-Savage-Abers-Cochran-2013}, among others).

Therefore, earthquake prevention strategies are necessary to mitigate induced seismicity in the energy sector \cite{b:Rubinstein-Mahani-2015, b:Zastrow-2019, b:Keranen-Savage-Abers-Cochran-2013}. Traffic light systems, cyclic stimulation and fracture caging are the most widely used approaches for earthquake prevention \cite{b:Verdon-Bommer-2021,b:10.1093/gji/ggz058,b:doi.org/10.1029/2020GL090648,https://doi.org/10.1007/s00603-018-1467-4}. Nevertheless, to our knowledge, these methods rely on trial and error rather than a systematic control approach. They lack a mathematical foundation and cannot guarantee the avoidance of induced seismicity. Moreover, there is no proof that these methods cannot even trigger earthquakes of greater magnitudes than the ones they are supposed to mitigate \cite{b:Tzortzopoulos-Braun-Stefanou-2021}. The fundamental assumptions of traffic light systems and its effectiveness were also questioned in \cite{b:10.1007/s10950-020-09966-9,b:10.1785/0220180337,b:10.1093/gji/ggac416}.

More recently, significant progress has been made in controlling the earthquake instability of specific, well-defined, mature seismic faults \cite{b:Stefanou2019, b:https://doi.org/10.1029/2021JB023410, b:Gutierrez-Tzortzopoulos-Stefanou-Plestan-2022, b:Gutierrez-Orlov-Stefanou-Plestan-2023, b:Gutierrez-Stefanou-Plestan-2022, b:Gutierrez-Orlov-Plestan-Stefanou-VSS2022}. These studies have employed various control algorithms to stabilize the complex and uncertain nature of the underlying underactuated physical system. The designed controllers effectively stabilize the system and modify its natural response time. As a result, the energy dissipation within the system occurs at a significantly slower rate, orders of magnitude lower than that of the natural (uncontrolled) earthquake event. However, it is worth noting that these investigations did not consider factors such as the presence of multiple smaller faults, which are typically found in deep geothermal reservoirs, as well as fluid circulation constraints associated with energy production. Regarding the controllability, observability and parameter identification of geological reservoirs, we refer to \cite{van_doren_controllability_2013,zandvliet_controllability_2008,van_doren_parameter_2011}.

In \cite{b:Schulze-Renner-2016}, the authors introduce a control strategy for a deep geothermal system. However, it is important to note that this study focused on a 1D diffusion system, the control design was implemented based on a discretized model (not accounting for possible spillover), and the output considered was the pressure over the reservoir, rather than the seismicity rate (SR). Therefore, this result presents many limitations to cope induced seismicity over real applications.
 
This study accounts for water injections into an underground reservoir using a simplified model of a 3D diffusion equation. By utilizing this Partial Differential Equation (PDE), a robust tracking control strategy is developed to regulate the SR, ensuring tracking to a desired reference. The primary control objective is to prevent induced seismicity while maintaining energy production. The designed control scheme demonstrates resilience against system uncertainties and unmodelled dynamics. Simulations of the process are conducted to validate the effectiveness of the control approach and show the potential for optimizing underground fluid circulation and thus energy production. Various simulation scenarios, considering different energy demands and constraints, are presented to provide a comprehensive understanding of the system's behaviour and the reliability of the control strategy. This research opens new perspectives for field applications at the kilometric scale based on real-time regulators and control theory.

The structure of this paper can be outlined as follows. In Section \ref{sec:motivation}, a motivating example of a simplified underground reservoir is presented showing how the SR increases when we inject fluids. Section \ref{sec:problem} introduces the underlying 3D diffusion model and shows the feedback control used for minimizing induced seismicity. To demonstrate the effectiveness of the proposed control strategy, simulations are conducted in Section \ref{sec:sim}, considering different scenarios of intermittent energy demand and production constraints. Finally, concluding remarks are provided in Section \ref{sec:conclusions}, summarizing the key findings of the study.

\section{EXAMPLE OF INDUCED SEISMICITY IN A RESERVOIR DUE TO FLUID INJECTIONS}
\label{sec:motivation}

\begin{figure*}[ht!]
  \centering 
  \includegraphics[width=13cm,keepaspectratio]{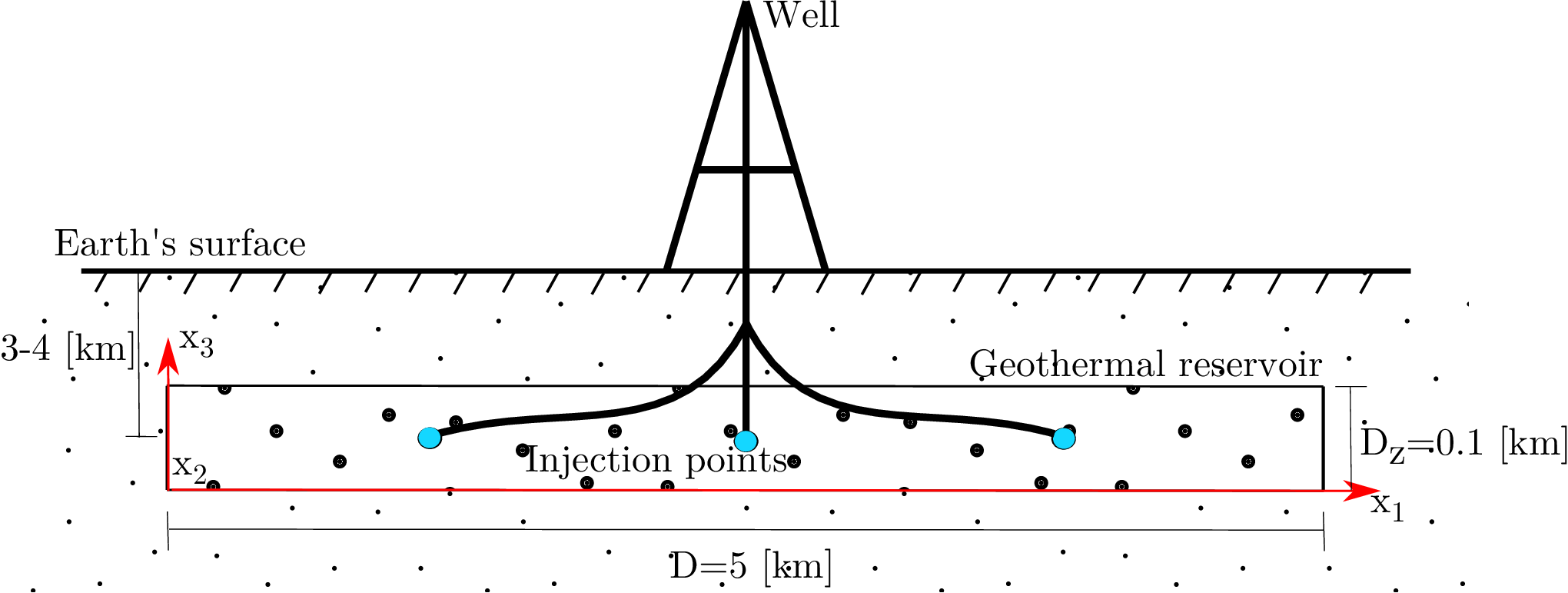}
  \caption{Underground reservoir diagram.}
  \label{fig:reservoir}
\end{figure*}

Consider a simplified underground reservoir at approximately $4$ [km] below the earth's surface, as depicted in Fig. \ref{fig:reservoir}. The reservoir is made of a porous rock which allows the circulation of fluids through its pores and cracks. In our example, the reservoir has a thickness of $\sim 100$ [m] and covers horizontally a square surface of dimensions $\sim 5 \text{ [km]}\times 5$ [km]. Wells are injecting and/or extracting fluids (\textit{e.g.} water) at different injection points in the reservoir, as shown in Fig. \ref{fig:reservoir}. For the sake of simplicity, injection of fluids will refer to both injection and fluid extraction from the reservoir.

Pumping fluids in-depth causes the circulation of fluids in the reservoir, which, in turn, causes the host porous rock to deform. The hydro-mechanical behaviour of the reservoir due to the injection of fluids at depth can be described by Biot's theory \cite{b:Biot-1941}. According to this theory, the diffusion of the fluid and the deformation of the host porous rock are coupled dynamic processes. However, if the injection rates are slow enough, with respect to the characteristic times of the system due to inertia, and if the volumetric strain rate of the host porous rock is negligible, then the diffusion of the fluid in the host rock due to fluid injections can be described by the following diffusion equation \cite{Zienkiewicz1980}
\begin{equation}
\begin{split}
  u_{t} &= -\frac{1}{\beta} \nabla q+ s,\\
\end{split}
\label{eq:diff1}
\end{equation}
where $u=u(x,t)$ is the change of the fluid pressure in the reservoir due to fluid injections, $x$ is the spatial coordinate, $t \geq 0$ is the time, $u_t$ denotes the partial derivative of $u$ with respect to time, $q=-\frac{k}{\eta} \nabla u$ is the change of the hydraulic flux and $s$ is a source/sink term representing the fluid injections. Furthermore, $k$ is the permeability of the host rock, $\eta$ is the dynamic viscosity of the fluid, and $\beta$ is the compressibility of the rock-fluid mixture. All these parameters are assumed constant in most of the following examples and, thus, they can define a simple expression for the hydraulic diffusivity of the system, $c_{hy}=\nicefrac{k}{\eta \beta}$. A special case where heterogeneities are present in $\beta$ and $c_{hy}$ is covered in Section \ref{sec:sim}. Finally, the reservoir has volume $V$.

We consider drained boundary conditions at the boundary of the reservoir, \textit{i.e.}, $u=0$ at $\partial V$. Furthermore, we assume point source terms, as the diameter of the wells is negligible compared to the size of the reservoir. In particular we set $s=\frac{1}{\beta}\mathcal{B}(x)Q(t)$, where $Q(t) \in \Re^{m}$, $Q(t)=[Q_{1}(t),...,Q_{m}(t)]^T$, are injection fluxes applied at the injection points, $(x^1,...,x^m)$, trough the coefficient $\mathcal{B}(x) \in \Re^{1 \times m}$, $\mathcal{B}(x) = [\delta(x-x^1),...,\delta(x-x^{m})]$. The terms $\delta(x-x^i)$ are Dirac's distributions and $m$ is the number of the wells in the reservoir. For the rigorous statement of the mathematical problem and its control we refer to Section \ref{sec:problem} and \ref{app:notation} to \ref{app:Control}.


It is nowadays well established that the injection of fluids in the earth's crust causes the creation of new and the reactivation of existing seismic faults, which are responsible for notable earthquakes (see for instance \cite{b:Rubinstein-Mahani-2015}, \cite{b:Keranen-Savage-Abers-Cochran-2013} and \cite{b:Zastrow-2019}). The physical mechanisms behind those human-induced seismic events is connected with the change of stresses in the host rock due to the injections, which intensify the loading and/or reduce the friction over existing or new discontinuities (faults). In other words, fluid injections increase the SR in a region, \textit{i.e.}, the number of earthquakes in a given time window. 

The seismicity rate, $R$, of a volume containing nucleation sources, \textit{i.e.} a region, depends on the average stress rate change, $\dot{\tau}$, over the same region according to the following expression
\begin{equation}
  \dot{R}=\frac{R}{t_a}\left( \frac{\dot{\tau}}{\dot{\tau}_0} -R\right),
  \label{eq:SR0}
\end{equation}
where $\dot{()}$ denotes the time derivative, $t_a$ is a characteristic decay time and $\dot{\tau}_0$ is the background stress change rate in the region, \textit{i.e.} the stress change rate due to various natural tectonic processes, and is considered to be constant. The above equation coincides with the one of Segall and Lu \cite{Segall2015} (see also \cite{Dieterich1994}), with the difference that here the SR is defined region-wise rather than point-wise. This choice results in a more convenient formulation as we mainly focus on averages over large volumes rather than point-wise measurements of the SR, which can be also singular due to point sources. Following Segall and Lu \cite{Segall2015} we assume also that the stress change rate is a linear function of the pore fluid pressure change rate, \textit{i.e.}, $\dot{\tau}=\dot{\tau}_0 + f \dot{u}$, where $\dot{u}$ is the average fluid pressure change rate over a given region of the reservoir and $f$ a (mobilized) friction coefficient. The latter linear hypothesis is justified on the basis of Coulomb friction over the fault planes and Terzaghi's effective stress principle \cite{b:Terzaghi-1943}.

It is worth emphasizing that, according to \cite{Dieterich1994}, the volume containing the nucleation sources must be small enough for all of them to experience uniform stress changes. Apparently, this is improbable during fluid injections in the Earth's crust, hence our simplifying assumption of averaging the Coulomb stress changes, $\dot{\tau}$, over a predefined region. The size of the averaging volumes (regions) could potentially be determined based on statistical arguments to accommodate the maximum expected induced earthquake magnitude (see for instance \cite{https://doi.org/10.1002/jgrb.50264}). However, these considerations extend beyond the scope of the current work (see also Section \ref{sec:conclusions} for more details about the limitations of the current approach). Finally, the point-wise SR model could be recovered at the limit of infinitesimal volumes. In this sense the region-wise SR approach used here is more general.

\begin{figure}[ht!]
  \centering 
  \includegraphics[width=6.5cm,keepaspectratio]{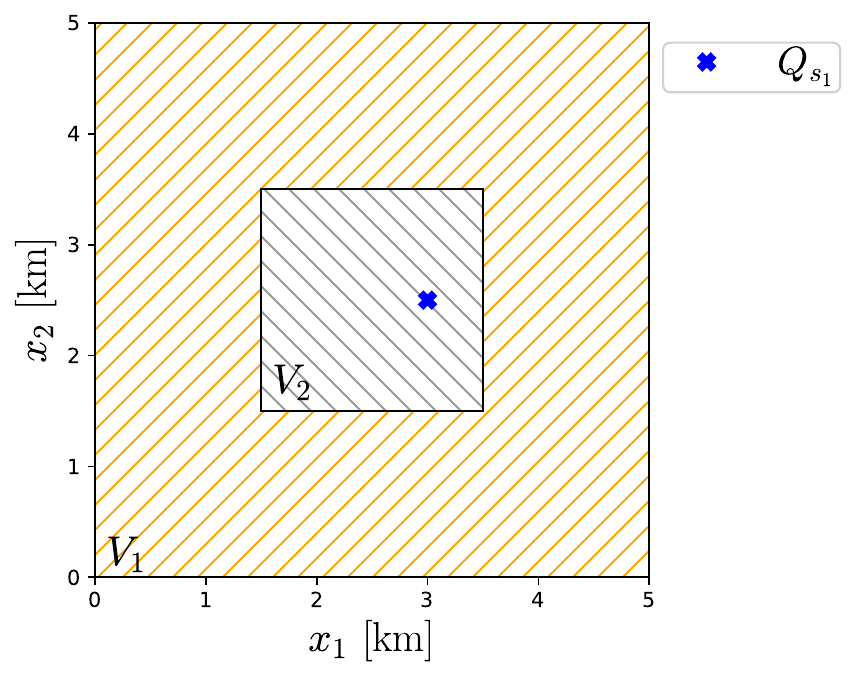}
  \caption{Regions $V_1$ and $V_2$ and location of the injection well with flux $Q_{s_1}$ inside of region $V_2$.}
  \label{fig:reservoir_no}
\end{figure}

In the absence of fluid injections, $\dot{\tau}=\dot{\tau}_0$ and, therefore, $R\rightarrow 1$. In this case, the SR of the region reduces to the natural one. If, on the contrary, fluids are injected into the reservoir, then $\dot {u}>0$  and consequently, $\dot{\tau}>\dot{\tau}_0$, leading to an increase of the SR ($\dot{R}>0$) over the region. To illustrate this mechanism, let us consider an injection of $Q=Q_{s_1}=32$ [m$^3$/hr] through a single injection well (see \cite{HARING2008469,KIM2022105098} for enhanced geothermal systems with similar injection rates). In this numerical example, we consider the parameters of Table \ref{tab:param}, we depth average Equation \ref{eq:diff1} as shown in \ref{app:average} and we integrate the resulting partial differential equation in time and space using a spectral decomposition method as explained in \ref{app:spectral}. We then calculate the SR over two distinct regions, one close to the injection point and one in the surroundings. Fig. \ref{fig:reservoir_no} shows the location of the regions and of the injection point. 

\begin{figure}[ht!]
  \centering 
  \includegraphics[width=6.8cm,keepaspectratio]{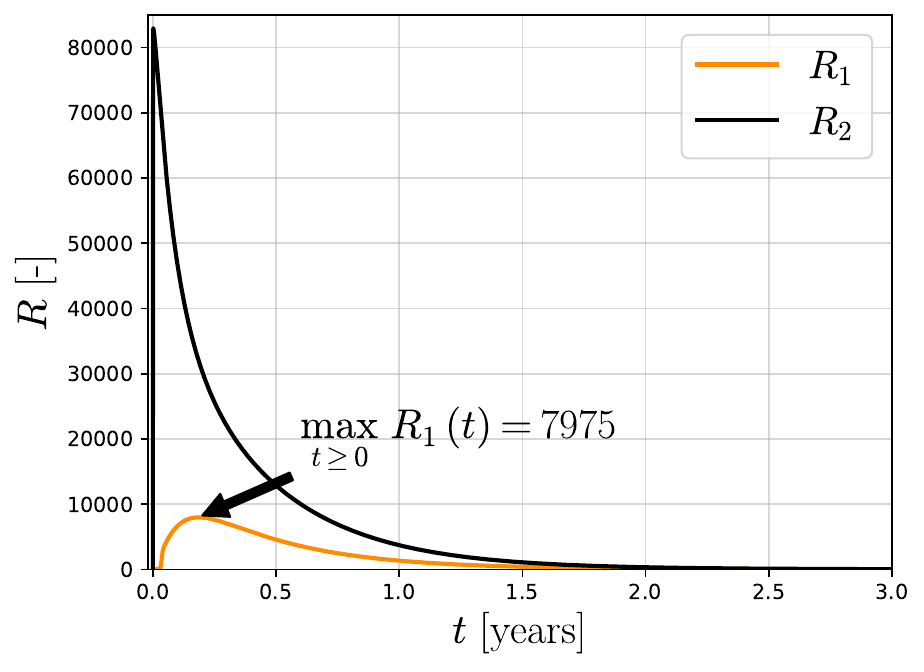}
  \caption{Seismicity rate in both regions, $V_1,V_2$ with constant injection rate, $Q_{s_1}$. $7975$ more earthquakes of a given magnitude in a given time window are expected over the outer region of the reservoir due to the constant fluid injection.}
  \label{fig:SR_no}
\end{figure}

\begin{figure*}[ht!]
  \centering 
  \includegraphics[width=5.1cm,keepaspectratio]{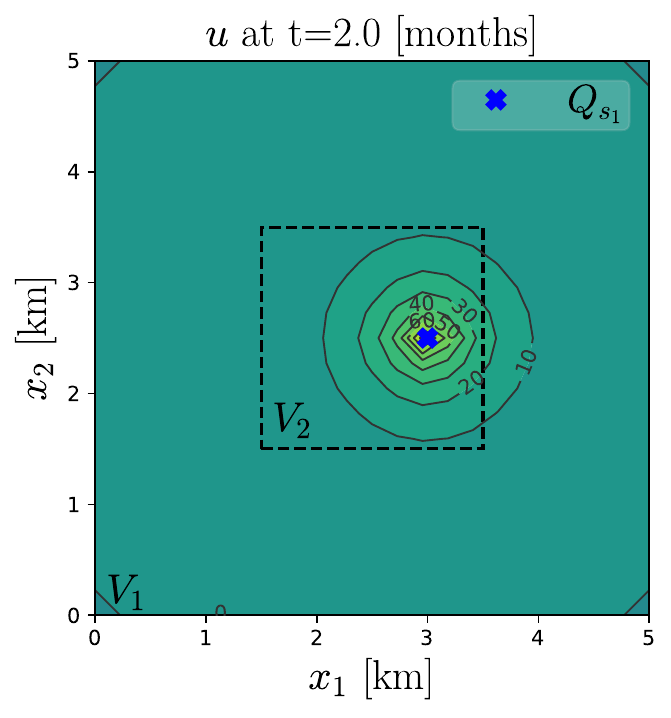}
  \includegraphics[width=5.1cm,keepaspectratio]{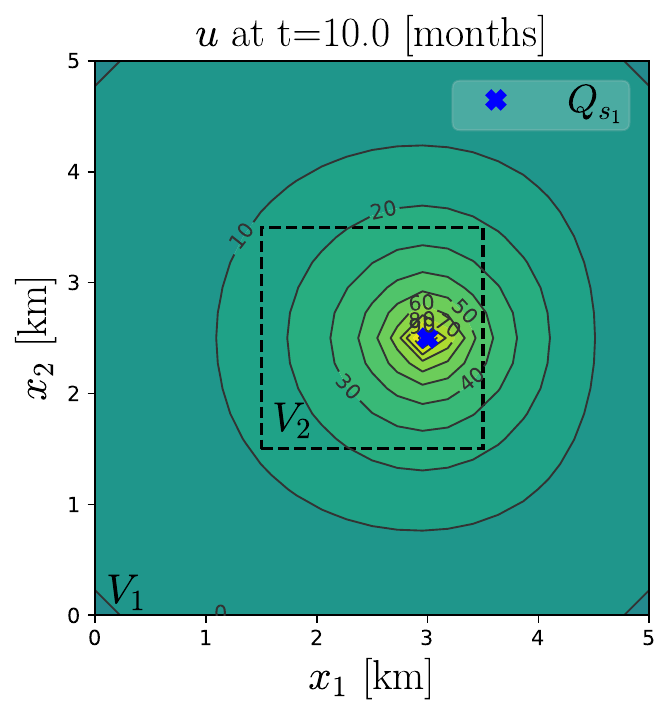}
  \includegraphics[width=6.0cm,keepaspectratio]{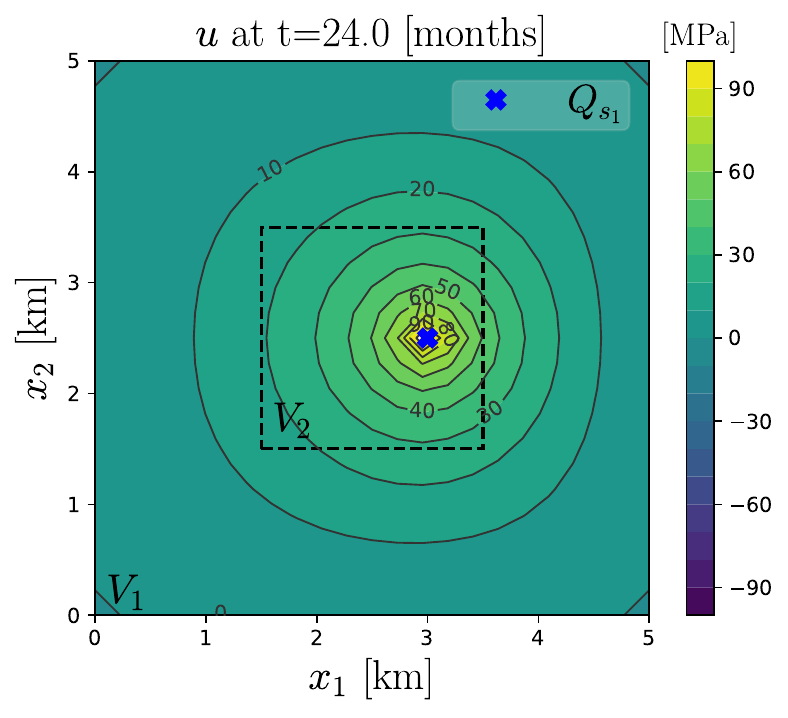}
  \caption{Solution, $u(x,t)$, of the pressure's reservoir at different times, with constant injection rate, $Q_{s_1}$. The solution presents high pressure profiles in wide areas next to the point of injection.}
  \label{fig:u_no}
\end{figure*}

The SR in both regions as a function of time is shown in fig. \ref{fig:SR_no}. We observe that the maximum SR over $V_1$ is equal to $R_1=7975$, which means that $7975$ more earthquakes of a given magnitude in a given time window are expected over region $V_1$, than without injections. The seismicity is even higher near the injection well (see region $V_2$ in Fig. \ref{fig:SR_no}). Fig. \ref{fig:u_no} shows the evolution of the pressure over the reservoir through different times. The pressure experiences a gradual rise across extensive areas near the injection point, eventually stabilizing at approximately two years.

In the case of an Enhanced Geothermal System (EGS \cite{Cornet2019}), we would like to increase the permeability between two wells by creating a small network of cracks that would facilitate the circulation of fluids between the wells \cite{https://doi.org/10.1002/nag.2330}. The creation of those cracks would be associated with a localized microseismicity in the region encircling the wells. This microseismicity is welcome, provided that the overall SR over the larger region of the reservoir remains close to one. Therefore, in the control problem addressed in this work, we will set as control objective the controlled increase of the SR in a small region surrounding some wells (\textit{e.g.}, in region $V_2$, see Fig. \ref{fig:reservoir_no}), while keeping constant and equal to one the SR over the larger area of the reservoir (\textit{e.g.}, in region $V_1$, see Fig. \ref{fig:reservoir_no}). For this purpose, additional wells will be added in the reservoir, whose fluxes will be controlled by a specially designed controller. This controller will be robust to uncertainties of the system parameters and will achieve the aforementioned control objective under different production rates. 

\begin{table}
\begin{center}
\caption{Diffusion and Seismicity rate system parameters. Such parameters are consistent with real applications like \cite{b://doi.org/10.1029/2019JB019134,b://doi.org/10.1002/2015JB012060}.}
{\footnotesize
\begin{tabular}{|c|c|c|}
\hline 
Parameter & Description & Value and Units \\ 
\hline 
$c_{hy}$ & Hydraulic diffusivity & $3.6 \times 10^{-4}$ [km$^2$/hr] \\ 
\hline 
$D$ & Reservoir length & $5$ [km] \\ 
\hline 
$D_z$ & Reservoir depth & $0.1$ [km] \\ 
\hline 
$Q_s$ & Static flux & $32$ [m$^3$/hr] \\ 
\hline 
$\beta$ & Mixture compressibility & $1.2 \times 10 ^{-4}$ [1/MPa] \\ 
\hline 
$f$ & Friction coefficient & $0.5$ [-] \\ 
\hline 
$\dot{\tau}_0$ & Background stressing rate & $1 \times 10 ^{-6}$ [MPa/hr] \\ 
\hline 
$t_a$ & Characteristic decay time & $500100$ [hr] \\ 
\hline 
\end{tabular}} 
\label{tab:param}
\end{center}
\end{table}


\section{PROBLEM STATEMENT AND CONTROL DESIGN}
\label{sec:problem}

The diffusion system \eqref{eq:diff1} can be written as follows
\begin{equation}
\begin{split}
  u_{t}(x,t) &= c_{hy} \nabla^2 u(x,t)+ \frac{1}{\beta}\left[\mathcal{B}_s(x) Q_s(t)+ \mathcal{B}_c(x) Q_c(t) \right],\\
  u(x,t) &= 0 \quad \forall \quad x \in S,\\
\end{split}
\label{eq:diff}
\end{equation}
where $u(x,t)$ is the fluid pressure change evolving in the space $H^1(V)$, $x \in \Re^3$, $x=[x_1,x_2,x_3]^T$, is the space variable belonging to a bounded subset $V$ of the space $\Re^3$ with boundary $S=\partial V$, and $t \geq 0$ is the time variable. As mentioned above, $c_{hy}$ is the hydraulic diffusivity and $\beta$ is the compressibility of the rock-fluid mixture. $Q_s(t) \in \Re^{m_s}$, $Q_s(t)=[Q_{s_1}(t),...,Q_{s_{m_s}}(t)]^T$, are fixed (not controlled) fluxes applied at the injection points, $(x_s^1,...,x_s^{m_s})$, trough the coefficient $\mathcal{B}_s(x) \in \Re^{1 \times m_s}$, $\mathcal{B}_s(x) = [\delta(x-x_s^1),...,\delta(x-x_s^{m_s})]$, and $Q_c(t) \in \Re^{m_c}$, $Q_c(t)=[Q_{c_1}(t),...,Q_{c_{m_c}}(t)]^T$, are the controlled fluxes applied at the injection points, $(x_s^1,...,x_s^{m_c})$, trough the coefficient $\mathcal{B}_c(x) \in \Re^{1 \times m_c}$, $\mathcal{B}_c(x) = [\delta(x-x_c^1),...,\delta(x-x_c^{m_c})]$. Note that the number of original inputs, $m$, in system \eqref{eq:diff1} is equal to the sum of not controlled and controlled input of system \eqref{eq:diff}, \textit{i.e.}, $m=m_s+m_c$ ($Q(t)=\mathcal{B}_s(x) Q_s(t)+ \mathcal{B}_c(x) Q_c(t)$). Since the right-hand side of \eqref{eq:diff1} contains Dirac's distributions, the above boundary value problem is interpreted in the weak sense (see \ref{app:notation} for more details on the notation and \ref{app:weak} for the definition of weak solution).

As explained before, the SR in equation \eqref{eq:SR0} is defined region-wise. In this study, we will define the SR over $m_c$ regions, $V_i \subset V$, $i \in [1,m_c]$ of the underground reservoir as follows
\begin{equation}
\begin{split}
  \dot{h}_i &= \frac{f}{t_a \dot{\tau}_0 V_i}\int_{V_i} u_t(x,t) \, dV - \frac{1}{t_a}(e^{h_i}-1), \quad i \in [1,m_c], 
\end{split}
  \label{eq:SR}
\end{equation}
where the change of variables
\begin{equation}
  R_i = e^{h_i}, \quad i \in [1,m_c],
  \label{eq:change}
\end{equation}
has been used for ease of calculation. The objective of this work is to design the control input $Q_c$ driving the output $y \in \Re^{m_c}$ defined as
\begin{equation}
\begin{split}
  y &= [h_1,...,h_{m_c}]^T,
\end{split}
  \label{eq:output}
\end{equation}
of the underlying BVP \eqref{eq:diff}--\eqref{eq:SR} to desired references $r(t) \in \Re^{m_c}$, $r(t)=[r_1(t),...,r_{m_c}(t)]^T$. This is known as tracking. Note that solving such tracking problem results in solving the tracking for the SR system \eqref{eq:SR0} due to the change of variables \eqref{eq:change}, \textit{i.e.}, $R_i(t)$ will be forced to follow the desired reference $\bar{r}_i(t)=e^{r_i(t)}$ for $i \in [1,m_c]$.

For that purpose, let us define the error variable, $y_e \in \Re^{m_c}$, as follows
\begin{equation}
 y_e = y-r,
\label{eq:error}
\end{equation}
and the control $Q_c(t)$ given by
\begin{equation}
\begin{split}
  Q_c(t) &= B_0^{-1} \left(-K_1 \Sabs{y_e}^{\frac{1}{1-l}} + \nu + \dot{r}\right), \\
  \dot{\nu} &= -K_2 \Sabs{y_e}^{\frac{1+l}{1-l}},
\end{split}
\label{eq:Qsr}
\end{equation}
where $K_1 \in \Re^{m_c \times m_c}$, $K_2 \in \Re^{m_c \times m_c}$ are matrices to be designed, and $l \in [-1,0]$ is a freely chosen parameter \cite{b:9901971,b:Mathey-Moreno-2022}. The matrix $B_0 \in \Re^{m_c \times m_c}$ is a nominal matrix that depends on the system parameters (see equation \eqref{eq:bounds} in \ref{app:Control} for its definition). The function $\lceil y_e \rfloor^{\gamma}:=|y_e|^{\gamma}\sign(y_e)$ is applied element-wise and is determined for any $\gamma\in \mathbb{R}_{\geq 0}$ (see \ref{app:notation} for more details). Such control has different characteristics depending on the value of $l$. It has a discontinuous integral term when $l=-1$, \textit{i.e.}, $\lceil y_e \rfloor^{0}=\sign(y_e)$. When $l \in (-1,0]$ the control function is continuous and degenerates to a linear integral control when $l=0$. Note how the controller is designed with minimum information about the system \eqref{eq:errordyn2}, \textit{i.e.} with only the measurement of $y(t)$ and the knowledge of the nominal matrix $B_0$. 

The tracking result for the output \eqref{eq:SR}--\eqref{eq:output} is then in force:

Let system \eqref{eq:diff}--\eqref{eq:SR} be driven by the control \eqref{eq:Qsr} with $K_1>0$, $K_2>0$ and $B_0$ as in \eqref{eq:bounds}. Then, the error variable \eqref{eq:error} will tend to zero in finite-time if $l=[-1,0)$, or exponentially if $l=0$.

In other words, we force the SR to follow desired references to avoid induced seismicity over the underground reservoir. (See \ref{app:Control} for the mathematical derivation of the proof and further details of the control algorithm.)

\subsection{Energy demand and production constraints}
\label{sec:demand}

We will consider a new scenario where an additional number of flux restrictions, $m_r$, to the fluid injection of the controlled injection points is needed. In other words, we will impose the weighted sum of the injection rates of some of the wells to be equal to a time-variant, possibly intermittent production rate.

For this purpose, we will augment the vector of controlled injection points as $\bar{Q}_c(t) \in \Re^{m_c+m_r}$, $\bar{Q}_c(t)=[\bar{Q}_{c_1}(t),...,\bar{Q}_{c_{m_c+m_r}}(t)]^T$. Notice that the number of controlled injection points, $m_c$, has to be increased to $m_r+m_c$. This does not change the previous theoretical results as shown below.

The condition imposed over the control input, $\bar{Q}_c(t)$, is
\begin{equation}
    W \bar{Q}_c(t) = D(t),
    \label{eq:restriction}
\end{equation}
where $W \in \Re^{m_r \times (m_c+m_r)}$ is a full rank matrix whose elements represent the weighted participation of the well's fluxes for ensuring the demand $D(t) \in \Re^{m_r}$. In order to follow this, the new control input will be designed as
\begin{equation}
   \bar{Q}_c(t) = \overline{W} Q_c(t) + W^T (W W^T)^{-1}D(t),
   \label{eq:Qcr}
\end{equation}
where $Q_c(t)$ is the original control input designed as \eqref{eq:Qsr} and \begin{footnotesize}{$\overline{W} \in \Re^{(m_c+m_r)\times m_c}$}\end{footnotesize} is the null space of $W$. Note that if we replace \eqref{eq:Qcr} in \eqref{eq:restriction}, the demand over the controlled injection points will be strictly fulfilled at any time $t$. 

Control \eqref{eq:Qcr} will ensure the linear combination of the new controlled fluxes $\bar{Q}_c(t)$ to be equal to a predetermined flux $D(t)$, which we called demand, according to \eqref{eq:restriction}, while keeping the original output tracking result of the previous section. This can be of interest in geoenergy applications to cope with different types of energy demand and production constraints (see \ref{app:demand} for more details).

Note that by monitoring the SR in the regions of interest and using equations \eqref{eq:Qsr} and \eqref{eq:Qcr} it is theoretically possible to adjust the fluid flux of the wells in a reservoir and achieve the desired control objectives in terms of the SR, while achieving production constraints.

\section{SIMULATIONS}
\label{sec:sim}

In order to demonstrate our control approach, numerical simulations of \eqref{eq:diff} and \eqref{eq:SR} have been done in Python (scripts available upon request). Without loss of generality and for a simpler presentation of the results, we chose to depth average Equation \ref{eq:diff} as shown in \ref{app:average} and integrate the resulting two-dimensional partial differential equation in time and space using Runge-Kutta method RK23 \cite{bogacki_32_1989} and spectral decomposition method as explained in \ref{app:spectral}, respectively. The same parameters with the numerical simulations performed in section \ref{sec:motivation} were used (see Table \ref{tab:param}).

\begin{figure}[ht!]
  \centering 
  \includegraphics[width=6.5cm,keepaspectratio]{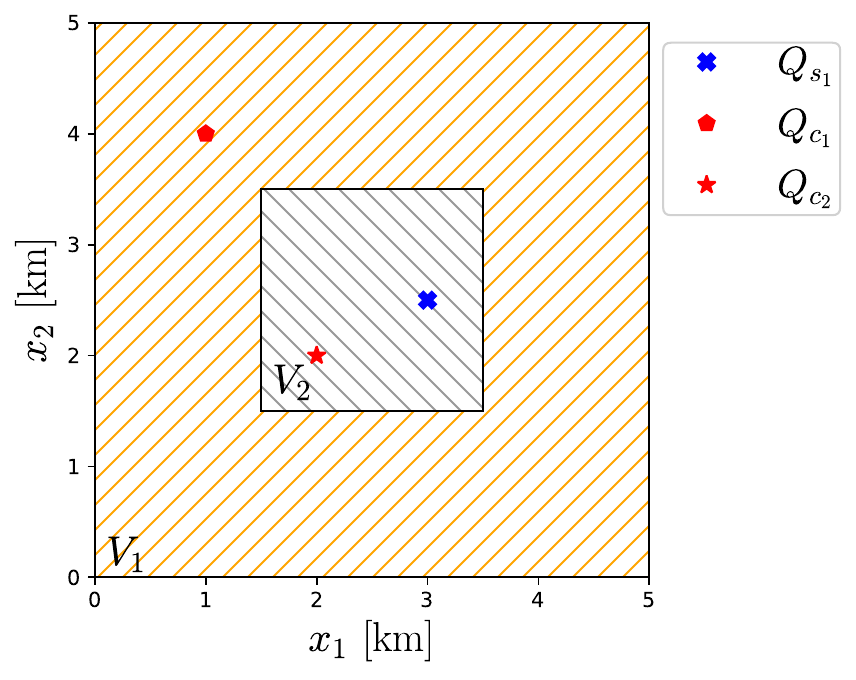}
  \includegraphics[width=6.5cm,keepaspectratio]{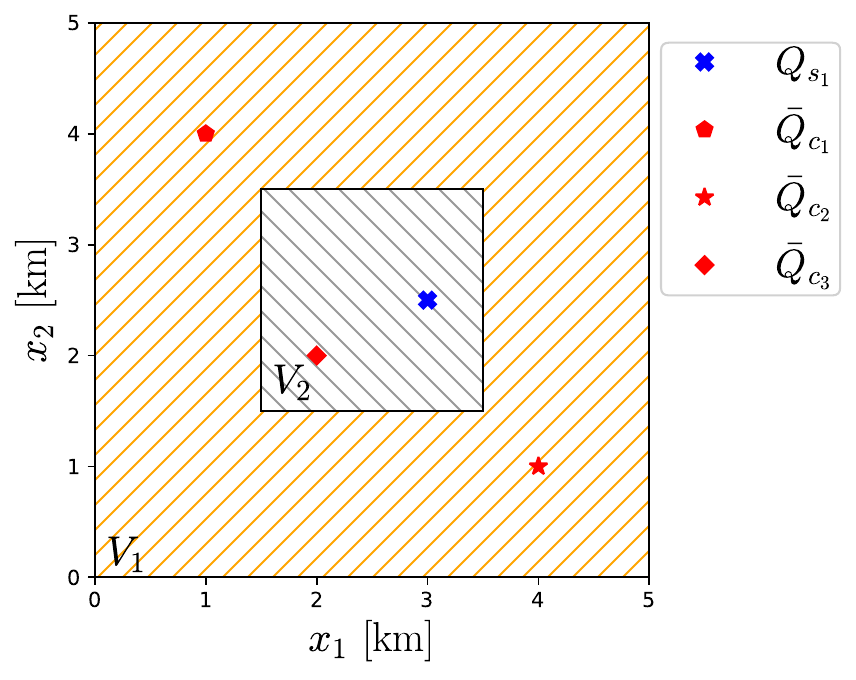}
  \caption{Regions $V_1$ and $V_2$ and location of the injection wells in the cases without demand (top) and with demand (bottom).}
  \label{fig:reservoir_control}
\end{figure}

Simulations were performed for three scenarios, i.e. one without any predetermined demand, one with constant demand and one with intermittent demand. In all scenarios we consider a fixed injection well with flux $Q_s(t)=Q_{s_1}(t)=32$ [m$^3$/hr] situated at the same location as the fixed injection well of the example presented in section \ref{sec:motivation}. Moreover, following the same example, we consider two different regions, $V_1$,$V_2$ over which we calculate the SR, $R_1$,$R_2$. Consequently, the number of outputs to be tracked is equal to two and, thus, at least two control inputs have to be designed ($Q_c(t)=[Q_{c_1}(t),Q_{c_{2}}(t)]^T$, $m_c=2$). In Fig. \ref{fig:reservoir_control} (top) we show the chosen location of the control wells. The initial conditions of the systems \eqref{eq:diff} and \eqref{eq:SR} were set as $u(x,0)=0$ and $h_1(0)=h_2(0)=0$ (\textit{i.e.}, $R_1(0)=R_2(0)=1$).

To design the control \eqref{eq:Qsr}, the matrices $B_s,B_c$ of system \eqref{eq:errordyn2} are needed (see \ref{app:Control} for more details), and they are defined as
\begin{equation}
  B_s = \frac{f}{t_a \dot{\tau}_0 \beta}\left[ 
  \begin{array}{c}
  0 \\
  \frac{1}{V_2} 
  \end{array}
  \right], \quad
  B_c =  \frac{f}{t_a \dot{\tau}_0 \beta} \left[ 
  \begin{array}{cc}
  \frac{1}{V_1} & 0 \\
  0 & \frac{1}{V_2} 
  \end{array}
  \right]. 
  \label{eq:matrices}
\end{equation}

For the scenarios with predetermined demand, we will apply a single flux restriction, \textit{i.e.}, $m_r=1$, with $W = [1, 1.01, 1]$ (see Section \ref{sec:demand} for more details). As a result, an addition control will be needed (\textit{i.e.} $m_c+m_r=3$), whose location is depicted in Fig. \ref{fig:reservoir_control} (bottom). Therefore, the matrices $\bar{B}_c,B_c$ of system \eqref{eq:errordyn3} become
\begin{equation}
\begin{split}
  \bar{B}_c &= \frac{f}{t_a \dot{\tau}_0 \beta}\left[ 
  \begin{array}{ccc}
  \frac{1}{V_1} & \frac{1}{V_1} & 0 \\
  0 & 0 & \frac{1}{V_2} 
  \end{array}
  \right], \\
  B_c &= \bar{B}_c \overline{W} = \frac{f}{t_a \dot{\tau}_0 \beta}\left[ 
  \begin{array}{cc}
  \frac{0.22}{V_1} & \frac{-0.78}{V_1} \\
  \frac{-0.21}{V_2} & \frac{0.79}{V_2} 
  \end{array}
  \right]
\end{split}
  \label{eq:matrices_demand}
\end{equation}
and the matrix $B_s$ is given in \eqref{eq:matrices}.  

Finally, in all scenarios, the reference $r(t)$ was selected as $r(t)=[r_1(t),r_2(t)]^T$, where $r_1(t)=0$ and $r_2(t)$ is a smooth function that reaches the final value of $\ln(5)$ in $1$ [month] (see Figs \ref{fig:SR1}, \ref{fig:SR2} and \ref{fig:SR3}). This reference was chosen so the SR on every region, $V_1,V_2$ to converge to the final values $1$ and $5$, respectively. This selection aims at forcing the SR in the extended region $V_1$ to be the same as the natural one. Regarding, region $V_2$ we opt for an increase of the SR in order to facilitate the circulation of fluids and improve the production of energy. 

\subsection{Scenario 1: SR tracking without demand}

In this scenario, the control \eqref{eq:Qsr} was implemented with a nominal matrix $B_0$ as
\begin{equation}
  B_0 = \frac{f_0}{t_{a_0} \dot{\tau}_{0_0} \beta_0}\left[ 
  \begin{array}{cc}
  \frac{1}{V_{1_0}} & 0 \\
  0 & \frac{1}{V_{2_0}}
  \end{array}
  \right],
  \label{eq:matrices2}
\end{equation}
where the subscript `0' corresponds to the nominal values of the system's parameters. In our case, we have chosen all the nominal values 10$\%$ higher than the real ones, \textit{e.g.}, $f_0=1.1 f$. 

The gain parameters of the control \eqref{eq:Qsr} were selected as
\begin{equation}
\begin{split}
  K_1 &= \left[ 
  \begin{array}{cc}
  1.5 \times 10^{-2} & 0 \\
  0 & 6.7 \times 10^{-2}
  \end{array}
  \right], \\
  K_2 &= \left[ 
  \begin{array}{cc}
  1.1 \times 10^{-4} & 0 \\
  0 & 2.2 \times 10^{-3}
  \end{array}
  \right], \quad l = -0.6
\end{split}
  \label{eq:matrices3}
\end{equation}

\begin{figure}[ht!]
  \centering 
  \includegraphics[width=6.8cm,keepaspectratio]{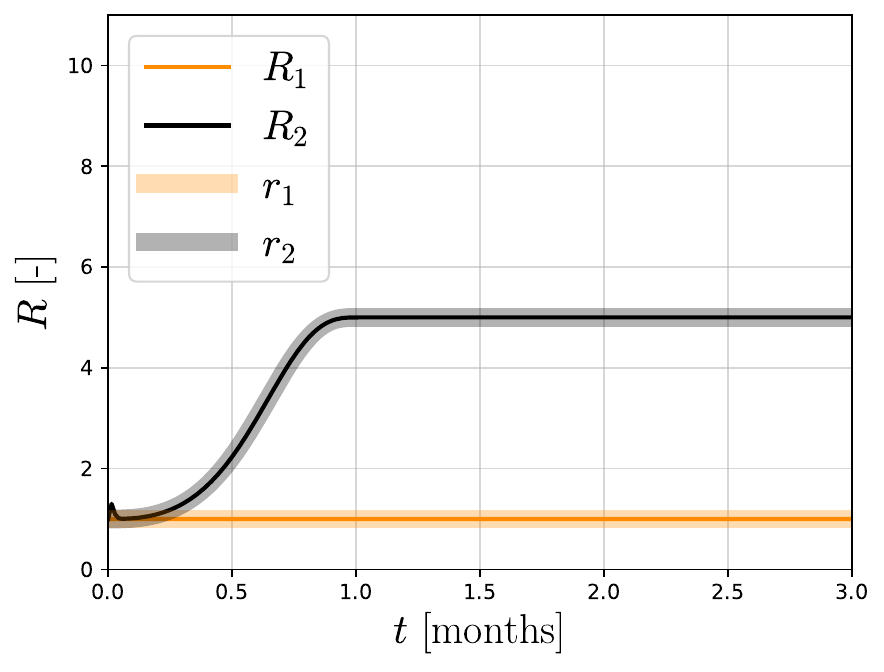}
  \caption{Seismicity rate in regions, $V_1,V_2$. The control strategy forces the SR to follow the desired references preventing induced seismicity.}
  \label{fig:SR1}
\end{figure}

\begin{figure*}[ht!]
  \centering 
  \includegraphics[width=5.1cm,keepaspectratio]{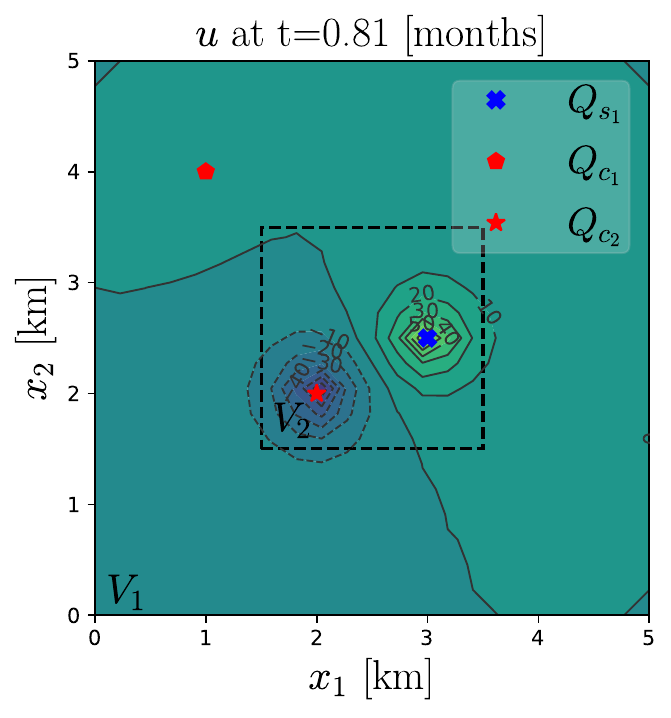}
  \includegraphics[width=5.1cm,keepaspectratio]{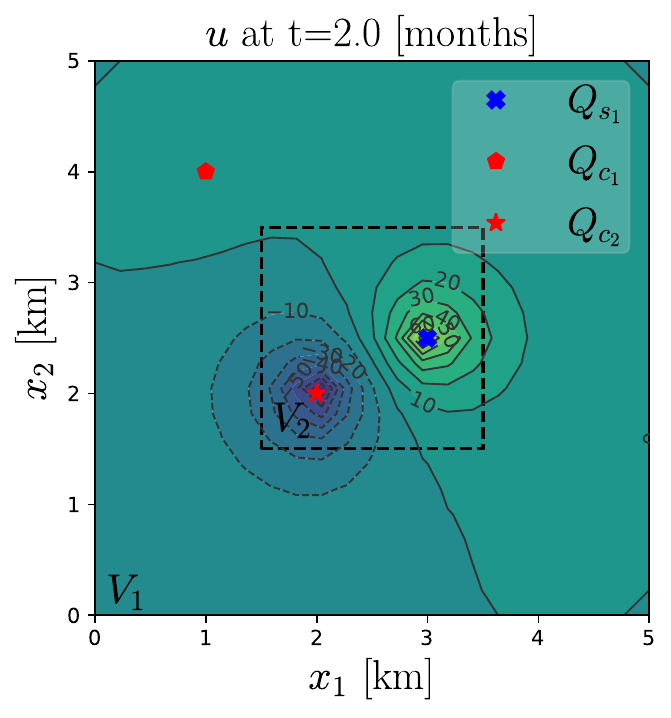}
  \includegraphics[width=6.0cm,keepaspectratio]{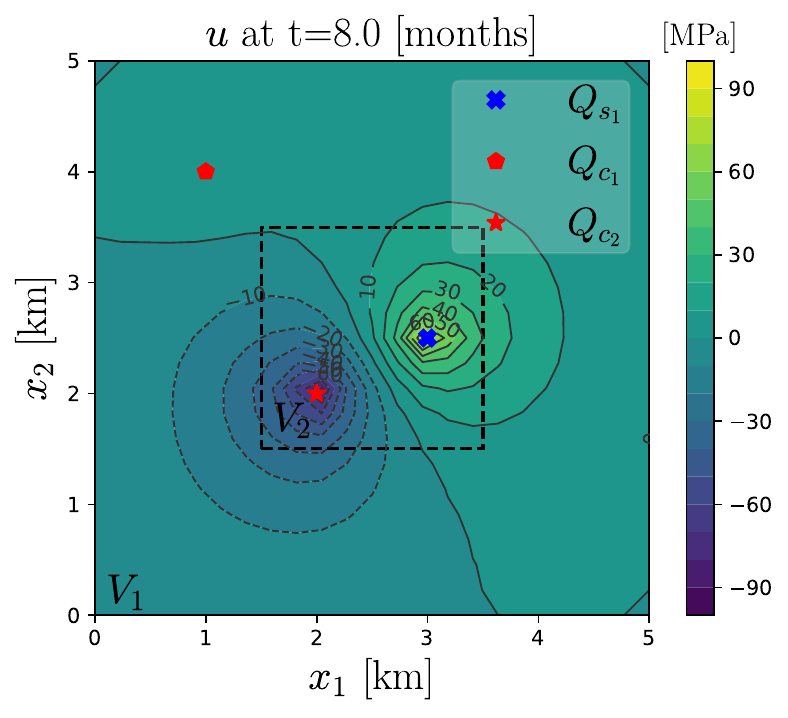}
  \caption{Fluid pressure distribution, $u(x,t)$, in the reservoir at different times. The control strategy prevents the propagation of high-pressure profiles around the static injection point.}
  \label{fig:u1}
\end{figure*}

\begin{figure}[ht!]
  \centering 
  \includegraphics[width=6.8cm,keepaspectratio]{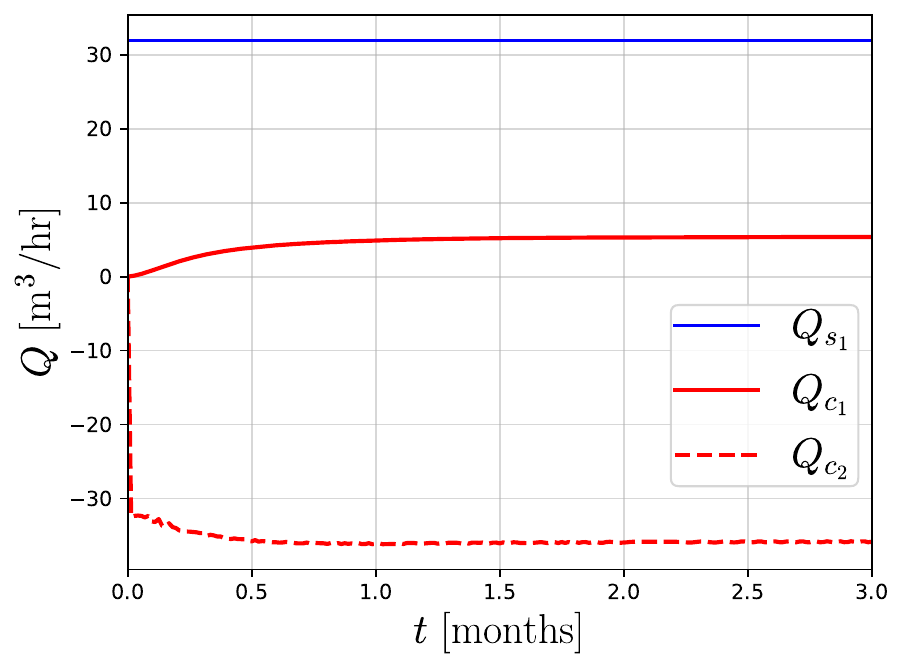}
  \caption{Static flux input $Q_{s_1}$ and controlled flux inputs $Q_{c_1},Q_{c_2}$.}
  \label{fig:Q1}
\end{figure}

The results are depicted in Figures \ref{fig:SR1} to \ref{fig:Q1}. In both regions, seismicity rates align with the specified constant references after approximately one month, achieving a steady state more rapidly than the uncontrolled system, as illustrated in Figure \ref{fig:SR_no}, which took around two years to reach a steady state. This robust performance is attributed to the control strategy, which effectively addresses uncertainties in the system. Specifically, the control uses only the nominal matrix $B_0$ and compensates for the remaining error dynamics. The generated control signals, presented in Figure \ref{fig:Q1}, exhibit continuous fluxes that can be used in practical pump actuators. Figure \ref{fig:u1} displays the pressure profile, $u(x,t)$, at different time points. In contrast to the scenario without control (refer to Figure \ref{fig:u_no}), the control strategy effectively prevents the propagation of high-pressure profiles around the static injection point.

\subsection{Scenario 2: SR tracking with constant demand}

\begin{figure}[ht!]
  \centering 
  \includegraphics[width=6.6cm,keepaspectratio]{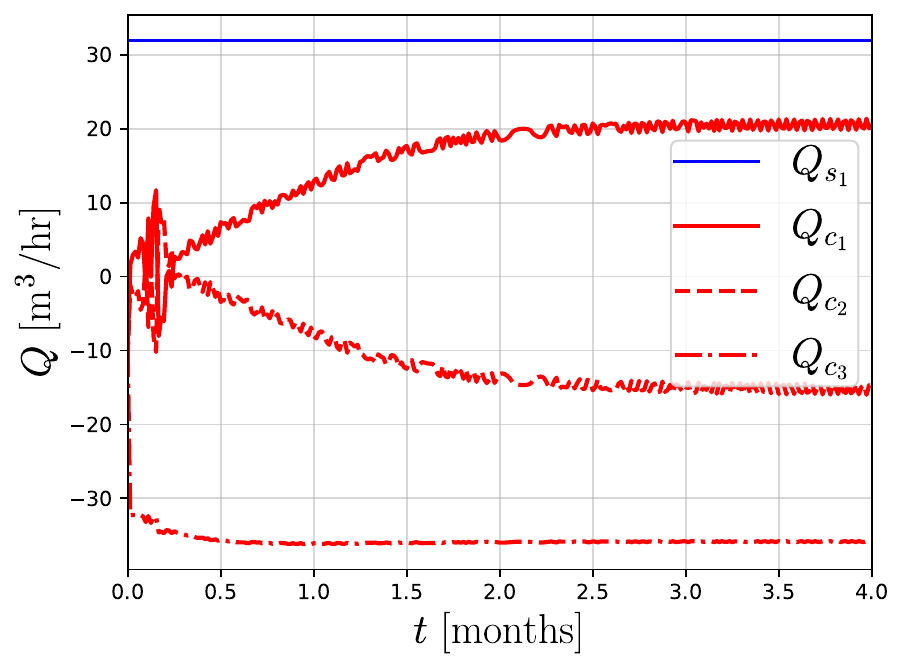}
  \includegraphics[width=6.6cm,keepaspectratio]{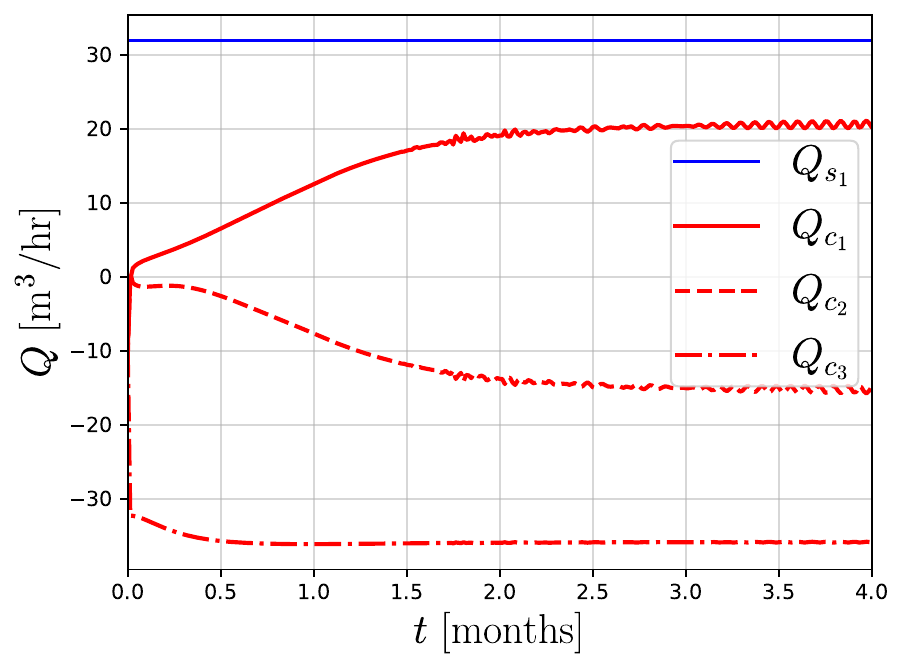}
  \includegraphics[width=6.6cm,keepaspectratio]{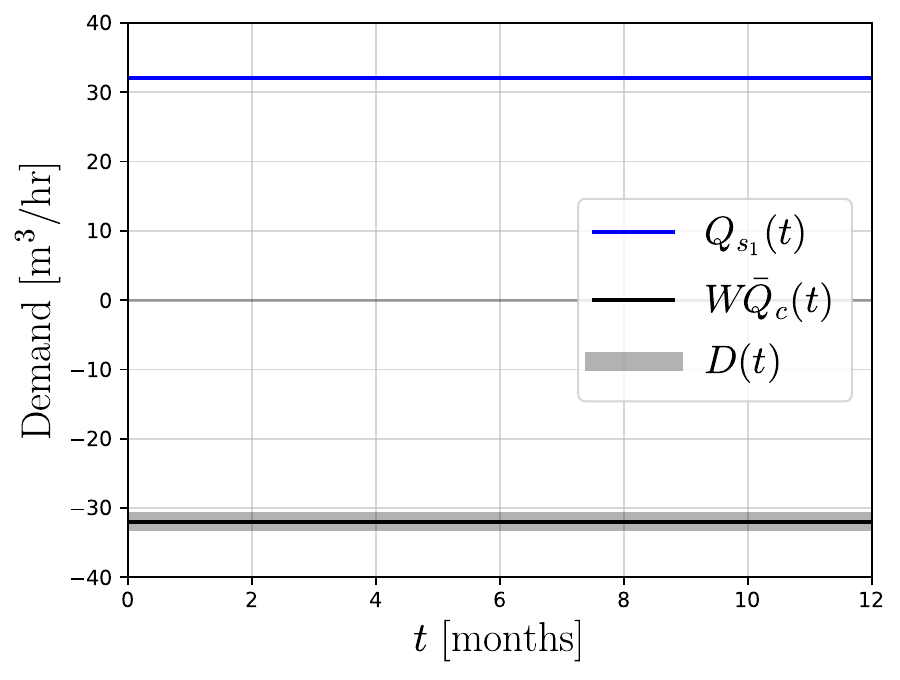}
  \caption{Injection flux of fixed well $Q_{s_1}$ and injection fluxes of the controlled wells $\bar{Q}_{c_1},\bar{Q}_{c_2},\bar{Q}_{c_3}$, under the constraint, $D(t)=W \bar{Q}_c(t)=-Q_{s_1}(t)$. Discontinuous case (top) and continuous case (center). The demand (bottom plot) is always ensured.}
  \label{fig:Q2}
\end{figure}

In this scenario we consider the demand to be equal to $D(t)=-Q_{s_1}=-32$ [m$^3$/hr] (see Fig. \ref{fig:Q2}). This is interesting in applications where the extracted fluid is re-injected into the reservoir. 

\begin{figure}[ht!]
  \centering 
  \includegraphics[width=6.8cm,keepaspectratio]{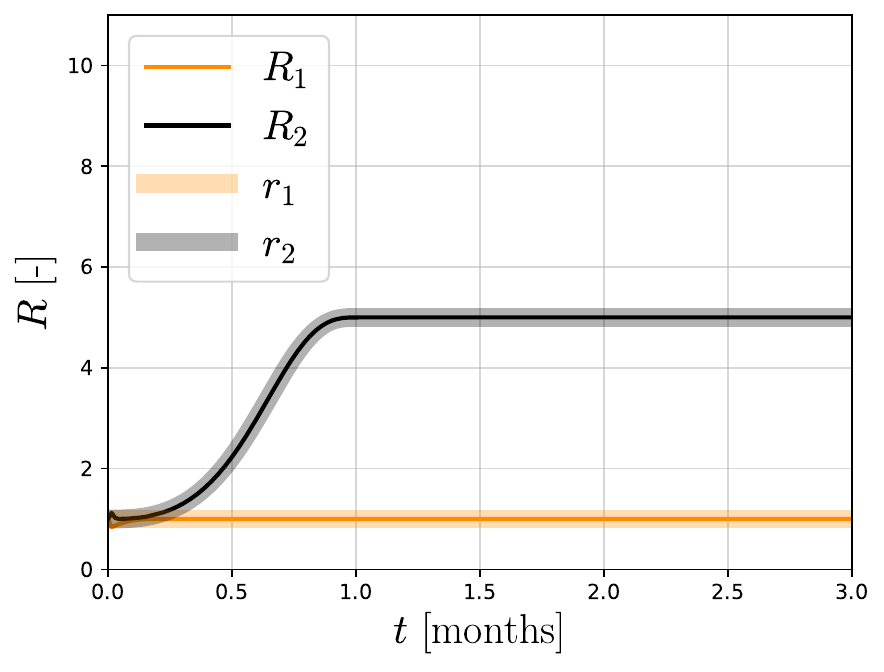}
  \caption{Seismicity rate in regions, $V_1,V_2$ under the constraint of constant demand. The control strategy forces the SR to follow the desired references while respecting the imposed energy demand.}
  \label{fig:SR2}
\end{figure}
\begin{figure*}[ht!]
  \centering 
  \includegraphics[width=5.1cm,keepaspectratio]{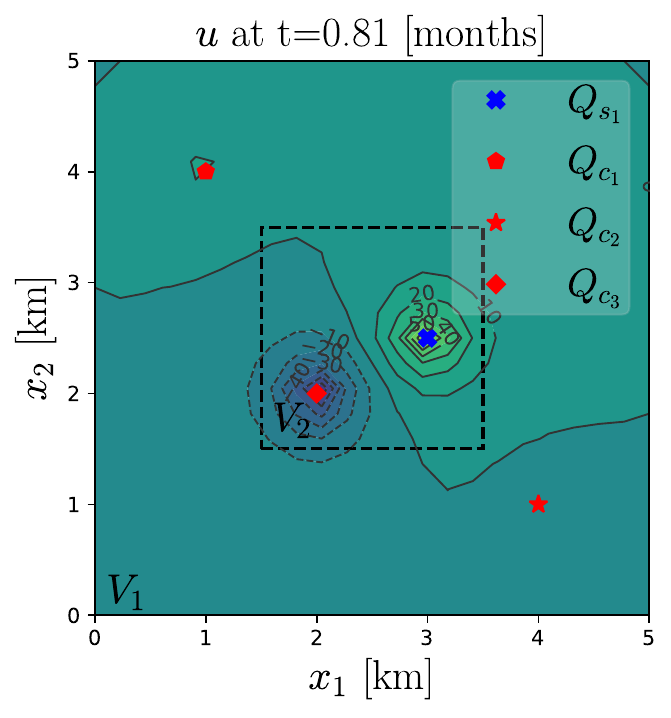}
  \includegraphics[width=5.1cm,keepaspectratio]{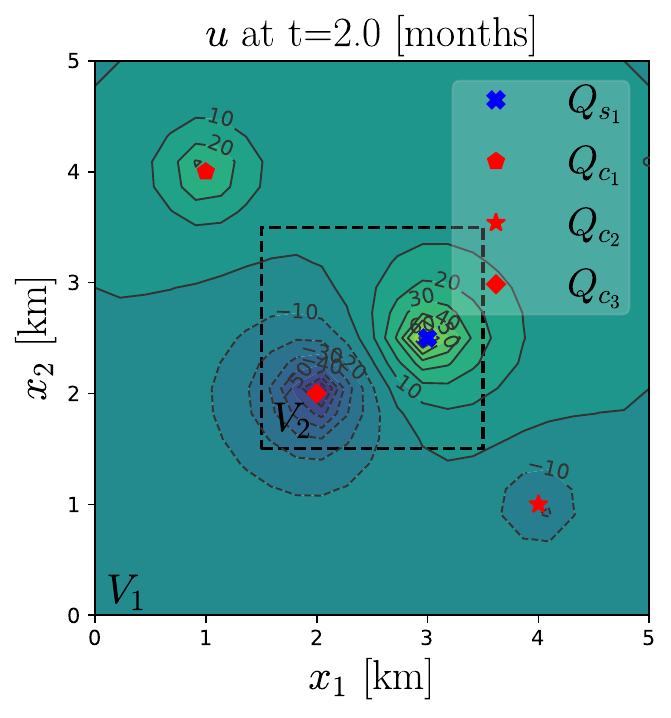}
  \includegraphics[width=6.0cm,keepaspectratio]{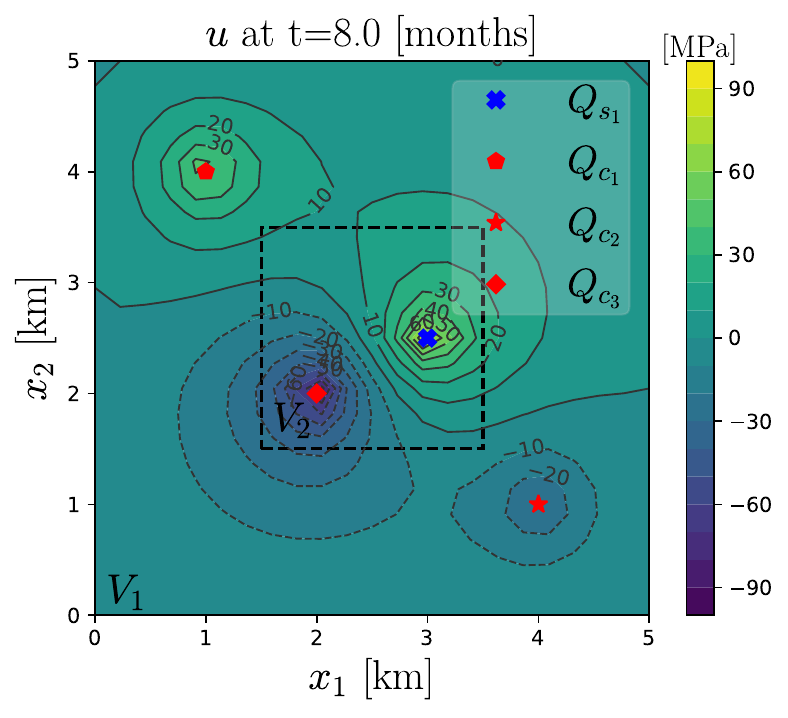}
  \caption{Pressure distribution, $u(x,t)$, in the reservoir at different times, under the constraint of constant demand. The control strategy regulates the SR across both regions, achieving stable low pressure solution profiles.}
  \label{fig:u2}
\end{figure*}

The control $\bar{Q}_c(t)$ was designed as \eqref{eq:Qsr},\eqref{eq:Qcr} with the nominal matrix $B_0$  
\begin{equation}
  B_0 = \frac{f_0}{t_{a_0} \dot{\tau}_{0_0} \beta_0} \left[ 
  \begin{array}{cc}
  \frac{0.22}{V_{1_0}} & \frac{-0.78}{V_{1_0}} \\
  \frac{-0.21}{V_{2_0}} & \frac{0.79}{V_{2_0}}
  \end{array}
  \right],
  \label{eq:matrices2_demand}
\end{equation}
where the subscript `0' corresponds to the nominal values of every system parameter. Again, we have chosen all the nominal values 10$\%$ larger than the real ones to test robustness. The control gains were selected as in \eqref{eq:matrices3} and the results are shown in Figs. \ref{fig:SR2}--\ref{fig:Q2}. Consistent with the earlier findings, the control effectively monitors the SR in two regions, but now under the influence of the imposed flux restriction \eqref{eq:restriction} on the control wells (refer to Fig \ref{fig:Q2}, bottom). The control strategy successfully regulates the seismicity rates across both regions, achieving a stable low pressure solution as in the previous case (see Figure \ref{fig:u2}).

\begin{figure}[ht!]
  \centering 
  \includegraphics[width=6.5cm,keepaspectratio]{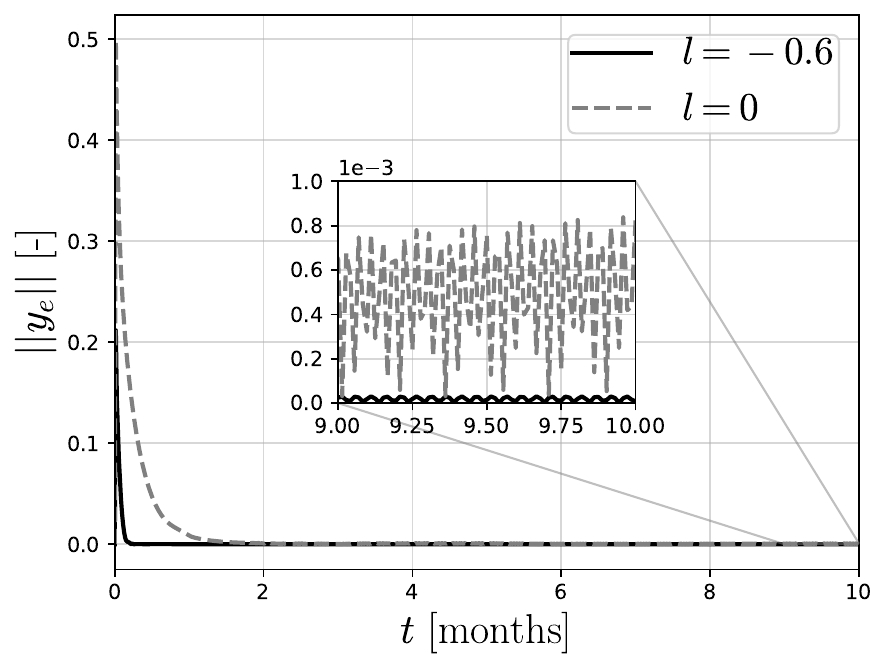}
  \caption{Norm of the tracking error, $y_e$, with constant demand using discontinuous control (continuous black line) and continuous control (dotted gray line).}
  \label{fig:e3}
\end{figure}

Figure \ref{fig:Q2} (top) shows fast-frequency oscillations (chattering) in the control signals due to the discontinuous nature of the employed controller. One can reduce such oscillations by choosing the linear case of the control, \textit{i.e.}, $l=0$ instead of $l=-0.6$. Indeed, in Fig. \ref{fig:Q2} (center) we observe the linear control signal generated for this case, which is smoother, but its convergence is slower and less precise. The latter is demonstrated in Fig. \ref{fig:e3}, where the norm of the tracking error is shown for both cases. Therefore, the selection of the control parameter, $l$, offers a trade-off between smoothness and precision, resulting in a great flexibility depending on the requirements of specific practical applications.

\subsection{Scenario 3: SR tracking with intermittent demand}

\begin{figure}[ht!]
  \centering 
  \includegraphics[width=6.8cm,keepaspectratio]{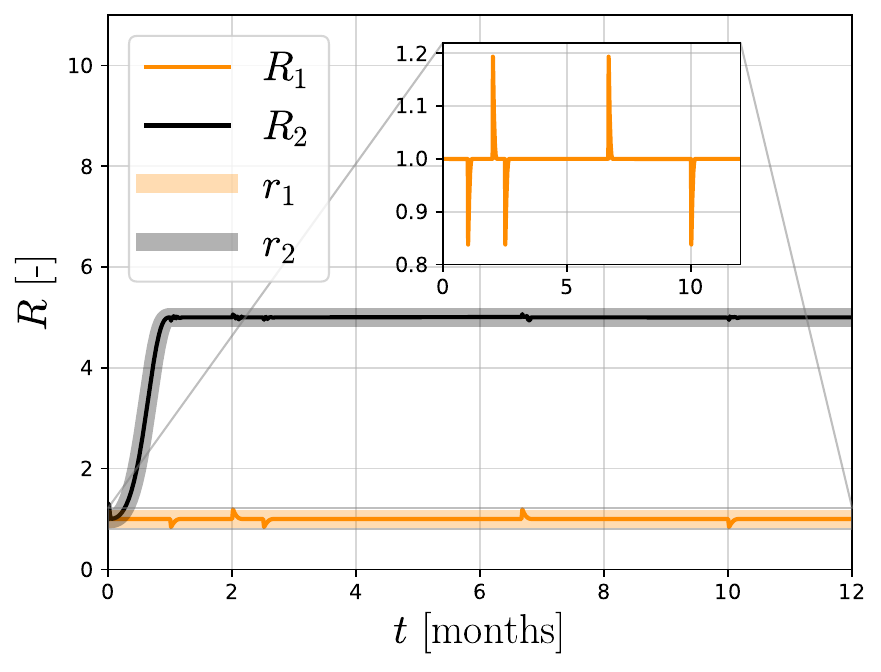}
  \caption{Seismicity rate in regions, $V_1,V_2$ under the constraint of intermittent demand. The control strategy forces the SR to follow the desired references while respecting the intermittent energy demand.}
  \label{fig:SR3}
\end{figure}

\begin{figure*}[ht!]
  \centering 
  \includegraphics[width=5.4cm,keepaspectratio]{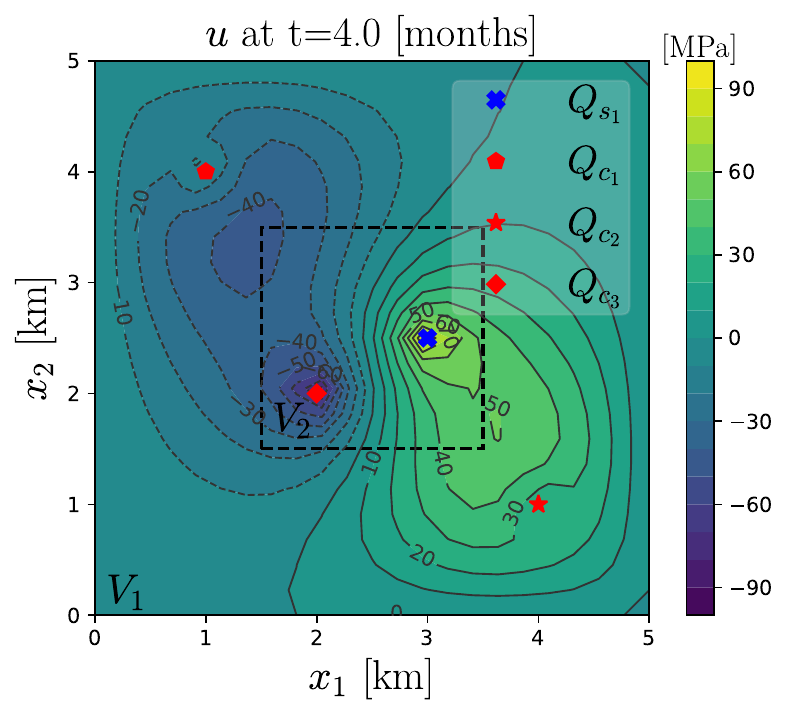}
  \includegraphics[width=5.4cm,keepaspectratio]{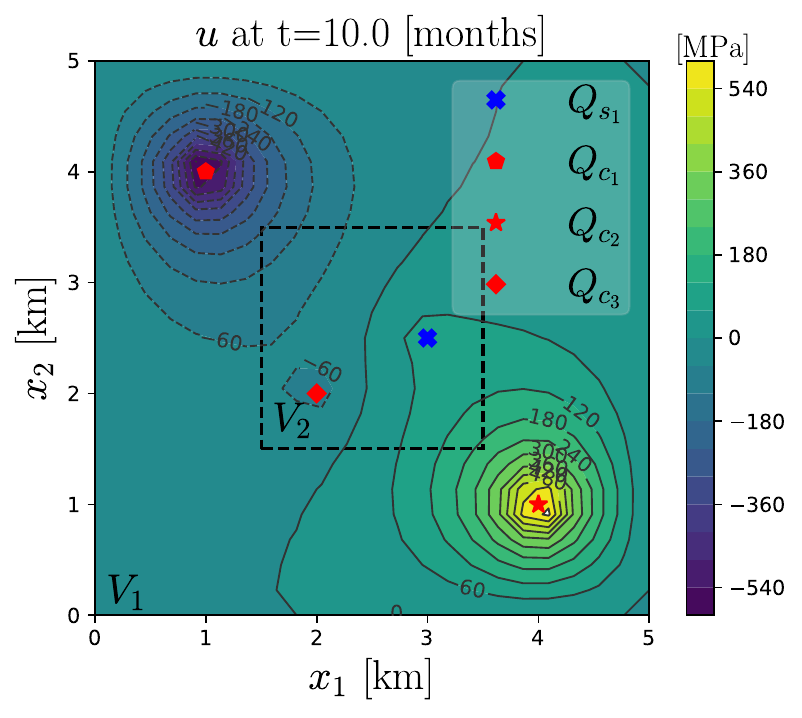}
  \includegraphics[width=5.4cm,keepaspectratio]{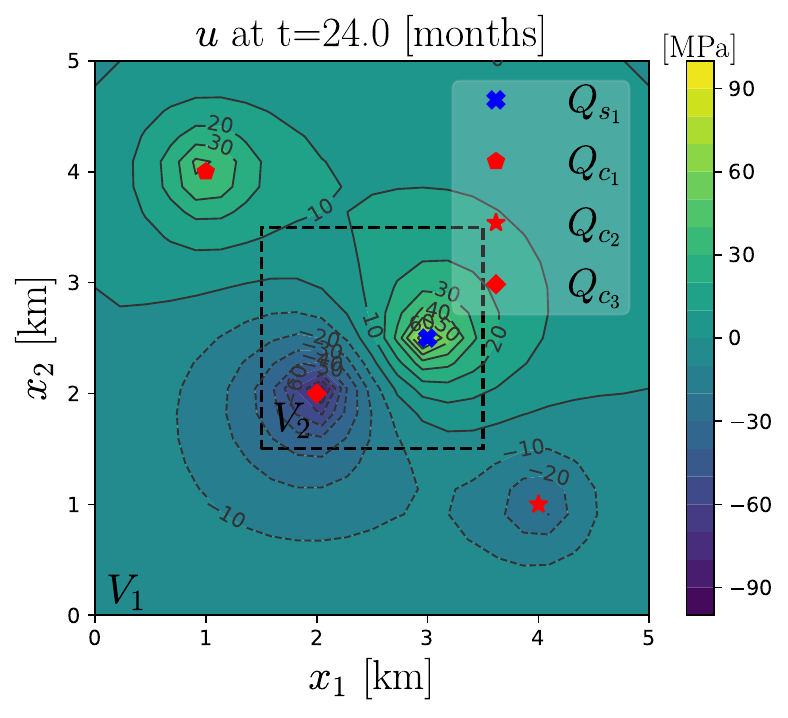}
  \caption{Fluid pressure distribution,  $u(x,t)$, in the reservoir at different times, under the constraint of intermittent demand. The abrupt shifts in demand result in higher pressure profiles compared to previous scenarios. Nevertheless, the control strategy ensures the steady state of the pressure solution after approximately 24 months.}
  \label{fig:u3}
\end{figure*}

\begin{figure}[ht!]
  \centering 
  \includegraphics[width=6.6cm,keepaspectratio]{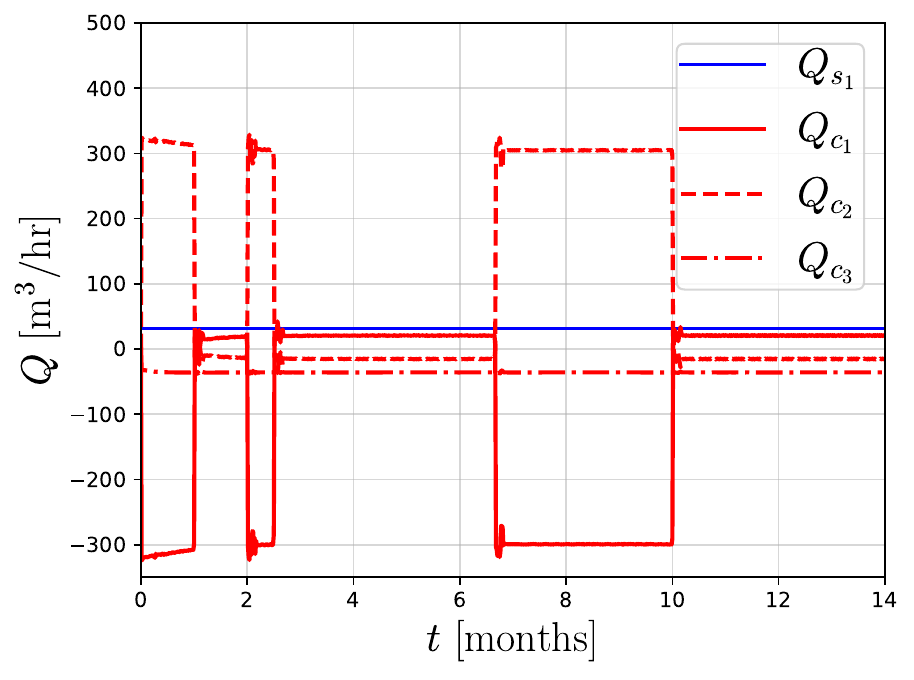}
  \includegraphics[width=6.6cm,keepaspectratio]{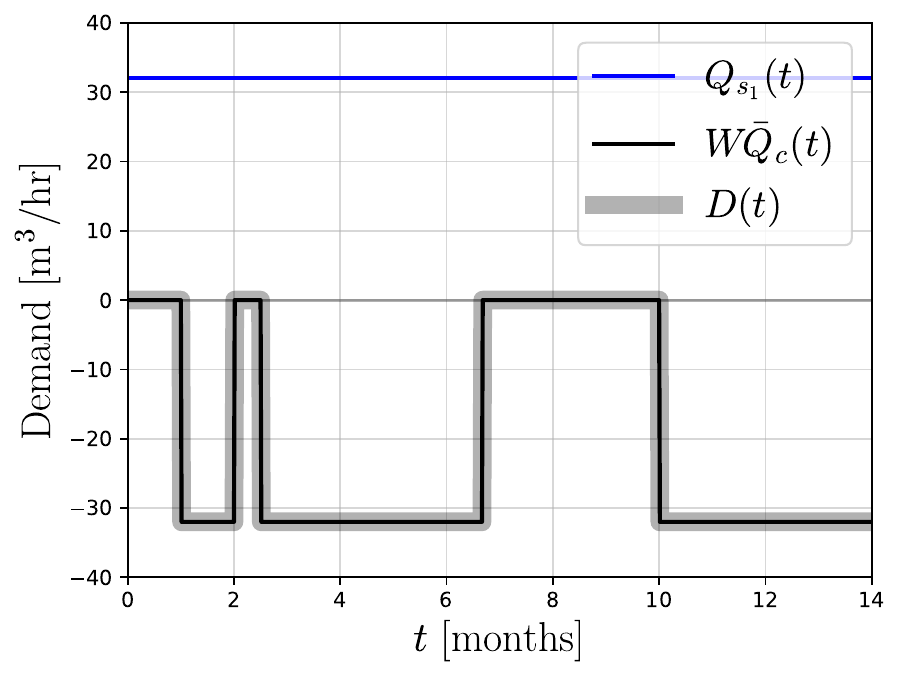}
  \caption{Injection flux $Q_{s_1}$ of the fixed well and injection fluxes $\bar{Q}_{c_1},\bar{Q}_{c_2},\bar{Q}_{c_3}$ of the control well under the constraint of intermittent demand. The demand (bottom) is strictly ensured again.}
  \label{fig:Q3}
\end{figure}

In order to test a more plausible demand scenario, we will apply the same control strategy as in the previous case, but using an intermittent demand (\textit{cf.} \cite{b://doi.org/10.1029/2019JB019134}), $D(t)$, as depicted in Fig. \ref{fig:Q3} (bottom). According to this injection plan, the demand varies abruptly between the flux of the fixed well and zero. The results are shown in Figs. \ref{fig:SR3}--\ref{fig:Q3}. Despite the intermittent nature of the demand, leading to numerous transients in the SR as depicted in Figure \ref{fig:SR3}, the generated control signals (Figure \ref{fig:Q3} (top)) swiftly restore the tracking performance for the SR. It is noteworthy that the control signals adhere strictly to the intermittent demand $D(t)$, as evident in Figure \ref{fig:Q3} (bottom). Additionally, the abrupt shifts in demand result in higher pressure profiles compared to previous scenarios, as illustrated in Figure \ref{fig:u3}. Nevertheless, the control strategy ensures the steady state of the pressure solution after approximately 24 months.

\subsection{Scenario 4: SR tracking with heterogeneities}

\begin{figure}[ht!]
  \centering 
  \includegraphics[width=6.0cm,keepaspectratio]{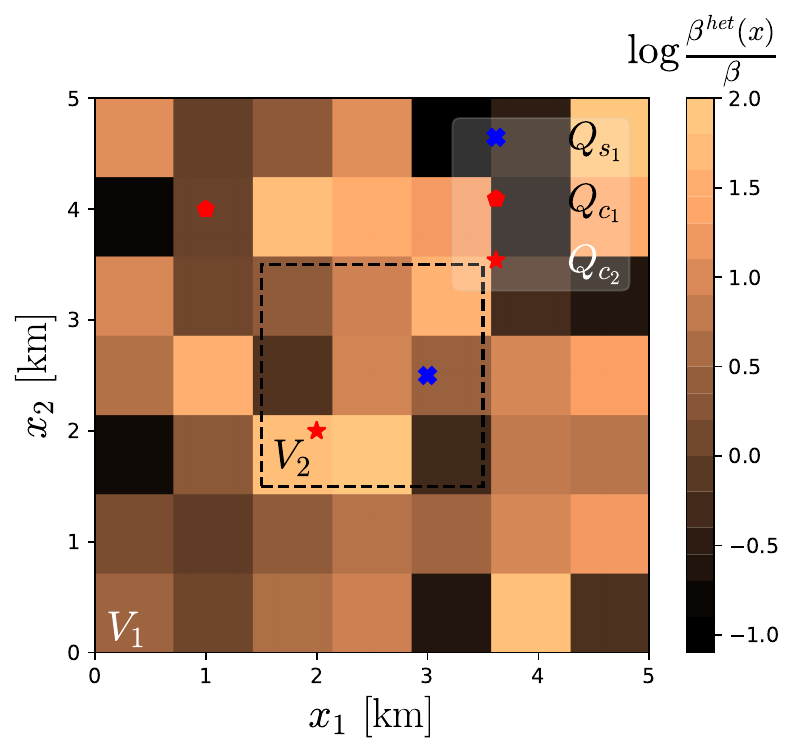}
  \includegraphics[width=6.0cm,keepaspectratio]{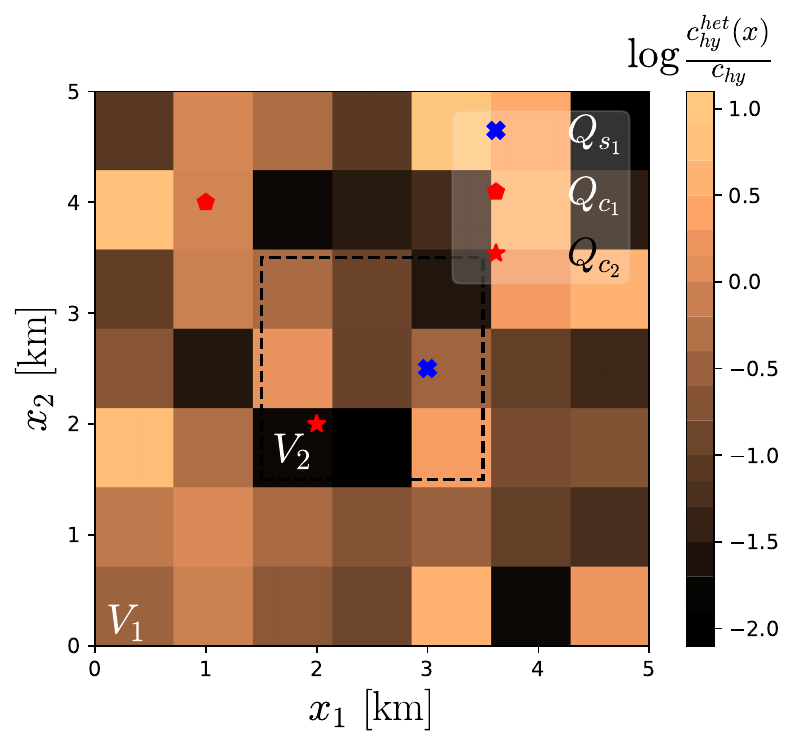}
  \caption{Heterogeneity of $\beta^{het}(x)$ and $c_{hy}^{het}(x)$ in the underground reservoir. The average weight of hydraulic diffusivity in the reservoir is $\nicefrac{c_{hy}^{het}(x)}{c_{hy}}=1.08$.}
  \label{fig:reservoir_het}
\end{figure}

As a final test to validate the robustness of the presented control strategy, we will introduce a more realistic scenario where there is presence of heterogeneities in the reservoir, more specifically in the hydraulic diffusivity and the mixture compressibility parameters, \textit{i.e.}, $c_{hy}=c_{hy}^{het}(x)$ and $\beta=\beta^{het}(x)$. Fig. \ref{fig:reservoir_het} illustrates the spatial variation of these parameters normalized by their constant values, as considered in the previous paragraphs. This variation extends to three orders of magnitude.

\begin{figure*}[ht!]
  \centering 
  \includegraphics[width=5.1cm,keepaspectratio]{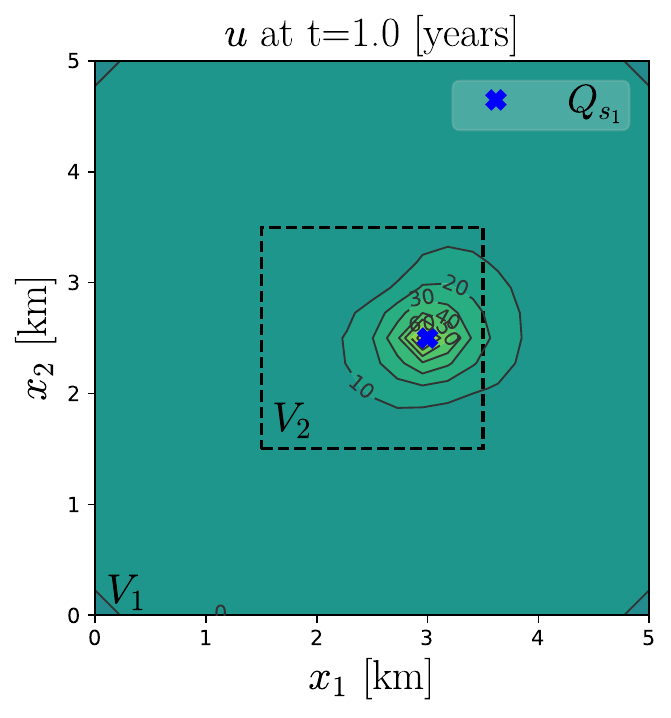}
  \includegraphics[width=5.1cm,keepaspectratio]{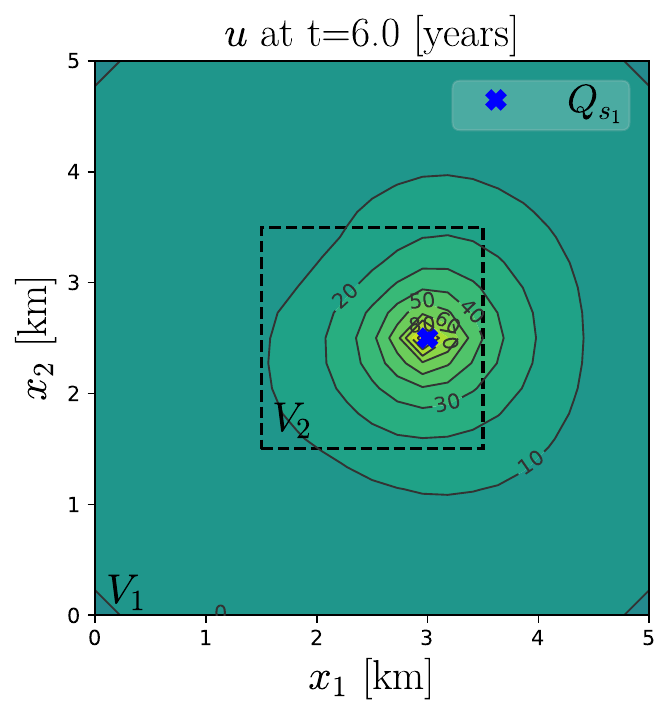}
  \includegraphics[width=6.0cm,keepaspectratio]{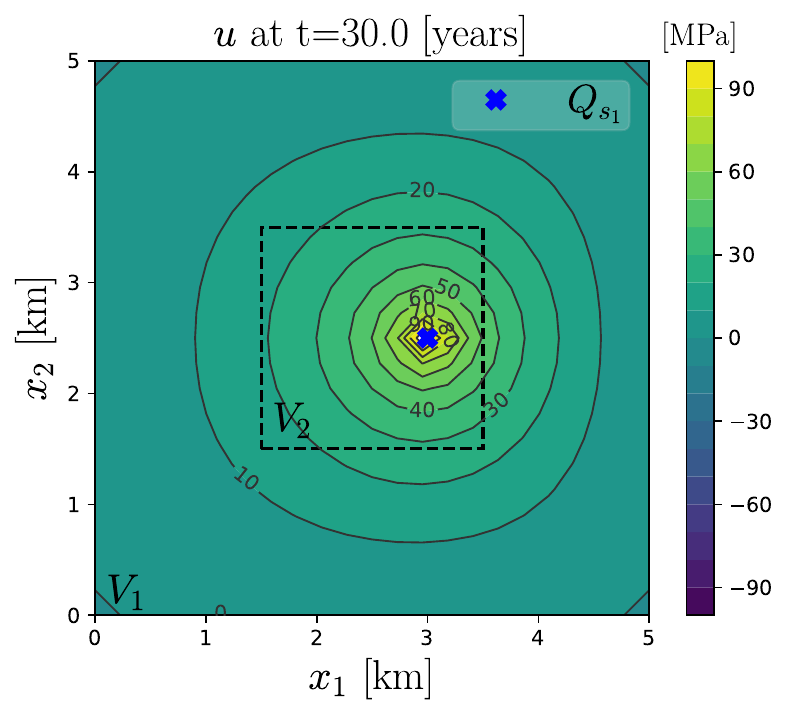}
  \caption{Pressure distribution, $u(x,t)$, at different times, under heterogeneities in the reservoir.}
  \label{fig:u4}
\end{figure*}

Fig. \ref{fig:u4} shows how the pressure evolves in the heterogeneous reservoir under the constant injection rate $Q_{s_1}(t)=32$ [m$^3$/hr] (no control). We observe that the contours are not circular convex anymore as in Fig. \ref{fig:u_no} due to heterogeneity. Furthermore, now the steady state is reached in a longer time (around 30 years in this simulation) due to the new distribution of $c_{hy}^{het}(x)$ and $\beta^{het}(x)$ across the reservoir.

\begin{figure*}[ht!]
  \centering 
  \includegraphics[width=5.1cm,keepaspectratio]{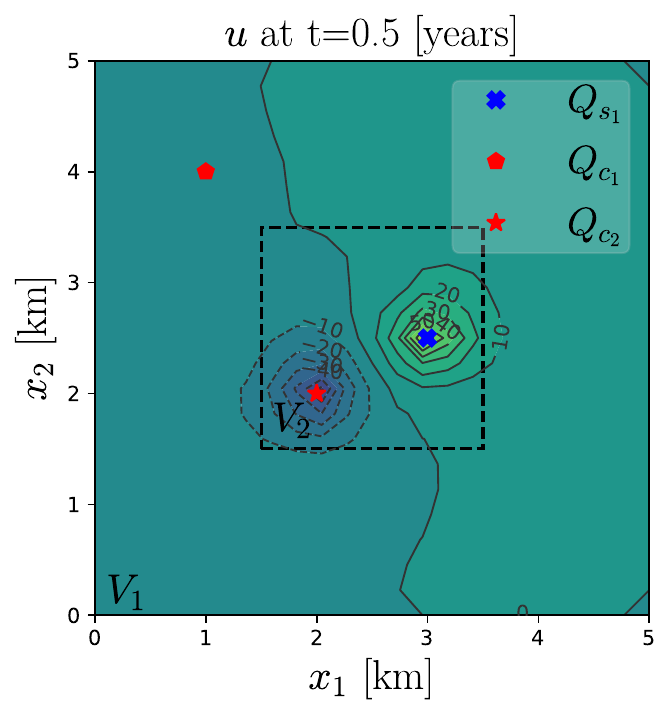}
  \includegraphics[width=5.1cm,keepaspectratio]{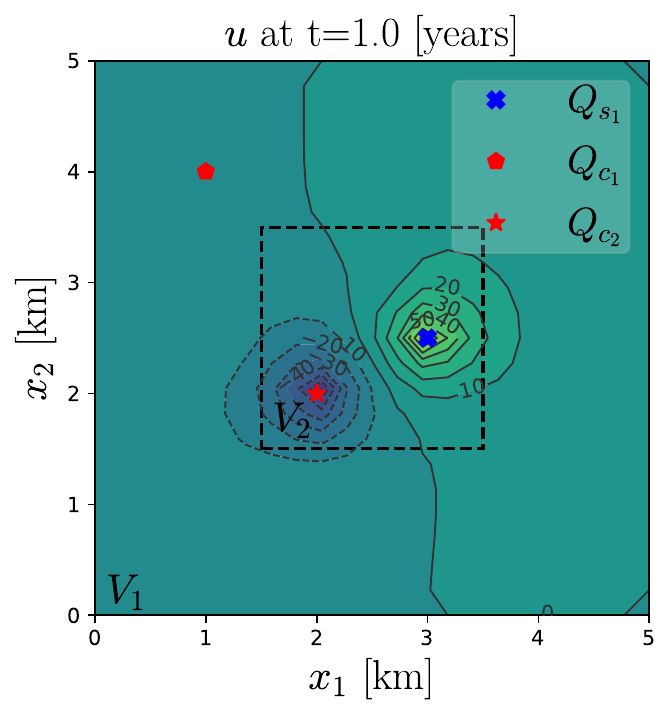}
  \includegraphics[width=6.0cm,keepaspectratio]{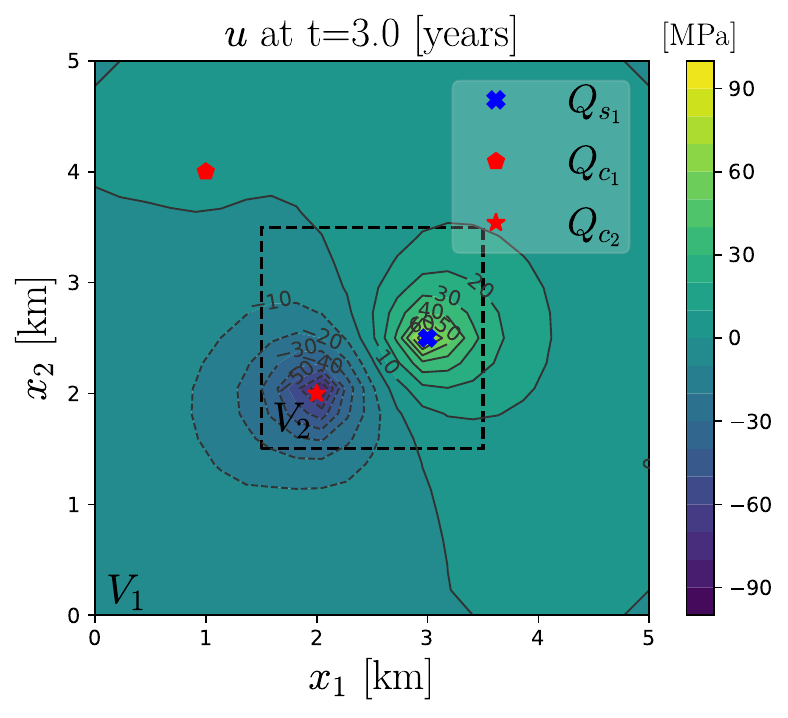}
  \caption{Pressure distribution, $u(x,t)$, at different times, under heterogeneities in the reservoir. The control strategy regulates the SR across both regions, achieving a stable low pressure solution after approximately 3 years.}
  \label{fig:u5}
\end{figure*}

\begin{figure}[ht!]
  \centering 
  \includegraphics[width=6.8cm,keepaspectratio]{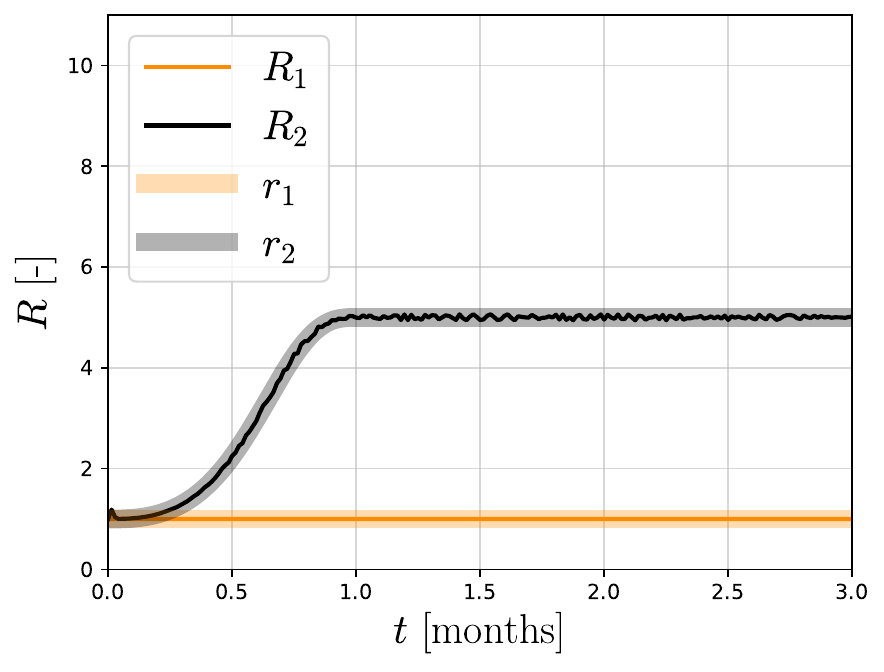}
  \caption{Seismicity rate in regions, $V_1,V_2$. The control strategy forces the SR to follow the desired references preventing induced seismicity.}
  \label{fig:SR4}
\end{figure}

\begin{figure}[ht!]
  \centering 
  \includegraphics[width=6.8cm,keepaspectratio]{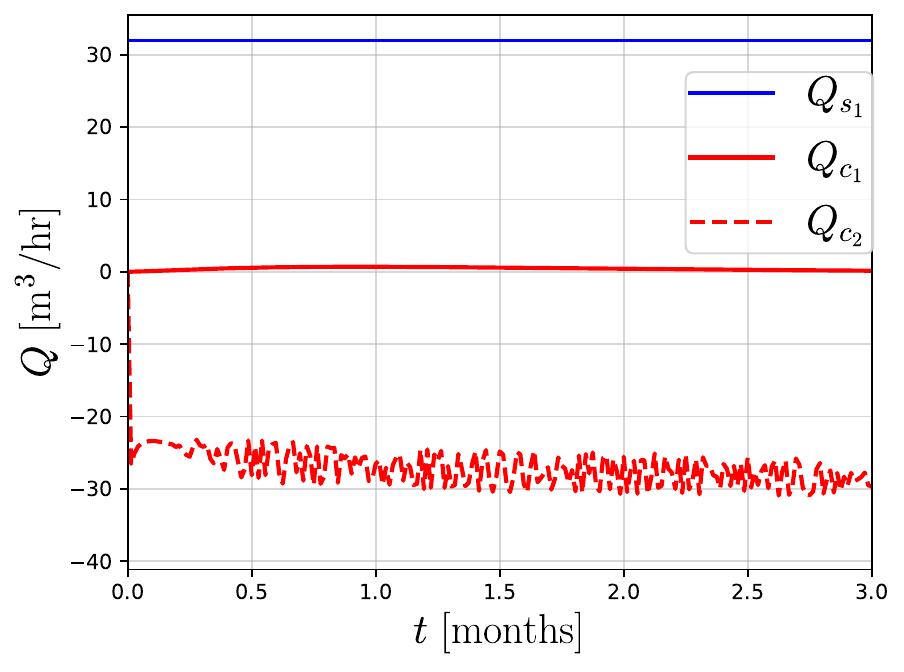}
  \caption{Static flux input $Q_{s_1}$ and controlled flux inputs $Q_{c_1},Q_{c_2}$.}
  \label{fig:Q4}
\end{figure}

The results of the simulations for the control case (Scenario 1) under heterogeneous parameters are depicted in Figs. \ref{fig:u5}--\ref{fig:Q4}. Although there are some oscillations in the control signal of the injection well $Q_{c_2}$ (see Fig. \ref{fig:Q4}) due to its location on the reservoir (see $\beta^{het}(x)$ value in Fig. \ref{fig:reservoir_het}), the control algorithm is still able to compensate such uncertainties and accomplish the tracking of the SR. The steady state of the pressure solution is achieved in a longer time compared to the case without heterogeneity (Fig. \ref{fig:u1}), but ten times faster than in the case without control (Fig. \ref{fig:u4}). 

\section{CONCLUSIONS AND LIMITATIONS}
\label{sec:conclusions}

A new control strategy for minimizing induced seismicity while assuring fluid circulation for energy production in underground reservoirs is designed in this paper. Contrary to existing approaches for induced seismicity mitigation due to fluid injections in the earth's crust, the proposed control strategy ensures the robust tracking of desired seismicity rates over different regions in geological reservoirs. For this purpose, a robust controller uses the measurement of averages of SR over different regions to generate continuous control signals for stabilizing the error dynamics, despite the presence of system uncertainties and unknown error dynamics. This is of great importance on this complicated system where it is always difficult to measure the real system parameters (\textit{e.g.}, diffusivity and compressibility), or there are errors in the sensing of physical quantities (\textit{e.g.}, SR and accelerometers).

A series of numerical simulations confirm the effectiveness of the presented theory over a simplified model of an underground reservoir under different scenarios. This provides a new direction for using robust control theory for this challenging application that involves an uncertain, underactuated, non-linear system of infinite dimensionality for mitigating induced seismicity while maximizing renewable energy production and storage.

However, assessing earthquake risk solely based on the seismicity rate may pose limitations, as earthquake magnitude holds greater significance than seismicity rate (\textit{c.f.}, Pohang EGS project \cite{KIM2022105098,https://doi.org/10.1029/2019JB018368} vs. Basel EGS project \cite{HARING2008469} earthquakes). The magnitude of expected earthquakes can be related to their frequency according to modified Richter-Gutenberg distributions as presented in \cite{https://doi.org/10.1002/jgrb.50264} and related works. The same holds for the maximum magnitude of the expected earthquakes, which may be connected to the size of the activated regions, $V_i$. However, the incorporation of this kind of statistical analysis exceeds the scope of this work.

Furthermore, the inclusion of fault discontinuities should be studied in real scenarios. Controlling multiple faults within a reservoir (\textit{cf.}, \cite{b:Boyet_2023}) remains a challenge due to complexity and faster spatio-temporal scales associated with poroelastodynamic phenomena activated by intermittent injections. Controlling this dynamics together with point-wise SR, instead of region-wise, nonlinear constraints for the fluxes and error in measurements, among others, is under investigation.

It is worth mentioning that the current approach has not yet been experimentally tested. A first step towards the experimental testing of the control of the slip of a single fault has been successful in the laboratory (see \cite{b:Gutierrez-Tzortzopoulos-Stefanou-Plestan-2022}), but this test seems rather adequate for intermittent injections and induced seismicity at the level of reservoirs. Moreover, field-scale testing is with no doubt necessary despite the methodological challenges related to repeatability and other methodological problems. Hence, this work is part of a more general effort for providing rigorous mathematical proofs and numerical simulations for exploring the extent up to which induced seismicity can be prevented in practice, while maximizing renewable energy production and storage.

\section*{Acknowledgement}
The authors would like to acknowledge the European Research Council's (ERC) support under the European Union’s Horizon 2020 research and innovation program (Grant agreement no. 757848 CoQuake and Grant agreement no. 101087771 INJECT). The second author would like to express his grateful thanks to Prof. Jean-Philippe Avouac for their fruitful discussions about human-induced seismicity and mitigation.


\bibliographystyle{elsarticle-num} 
\bibliography{Bibliografias}

\appendix

\section{Notation}
\label{app:notation}

We denote $\norm{\cdot}$ as the euclidean norm of the $n$-dimensional Euclidean space, $\Re^n$. If $y_e \in \Re$, the function $\lceil y_e \rfloor^{\gamma}:=|y_e|^{\gamma}\sign(y_e)$ is determined for any $\gamma\in \mathbb{R}_{\geq 0}$. If $y_e \in \Re^m$, the functions $\lceil y_e \rfloor^{\gamma}$ and $|y_e|^{\gamma}$ will be applied element-wise. 

Consider the Sobolev space, $H^1(V)$, of absolutely continuous scalar functions $u(x)$, $x \in \Re^3$, $x=[x_1,x_2,x_3]^T$, defined on a bounded subset $V$ of the space $\Re^3$ with boundary $S=\partial V$ as $H^1(V)=\left\{ u \mid u, \nabla u \in {\cal L}^2(V) \right\}$. Its $H^0$-norm is defined as $\norm{u(\cdot)}_{H^0(V)} =\sqrt{\int_V [u(\cdot)]^2\, dV}$. The time derivative is denoted by $u_t =\nicefrac{\partial u}{\partial t}$, the gradient as $\nabla u=\left[\nicefrac{\partial u}{\partial x_1},\nicefrac{\partial u}{\partial x_2},\nicefrac{\partial u}{\partial x_3}\right]$, and the Laplacian as $\nabla^2 u=\nicefrac{\partial^2 u}{\partial x_1^2}+\nicefrac{\partial^2 u}{\partial x_2^2}+\nicefrac{\partial^2 u}{\partial x_3^2}$. We define the Dirac's distribution, $\delta(x)$, as $\int_{V^*} \phi(x) \delta(x-x^*)\, dV=\phi(x^*)$, $\forall$ $x^* \in V$, $V^* \subset V$, on an arbitrary test function $\phi(x) \in H^1(V)$.

For their use, important inequalities are recalled:\\
\noindent \textbf{Poincar\'e's Inequality:} For $u(x)\in H^1(V)$ on a bounded subset $V$ of the space $\Re^3$ of zero-trace (\textit{i.e.}, $u(x,t) = 0$ for all $x \in \partial V$), the inequality
\begin{equation}
  \norm{u(x)}_{H^0(V)}^2 \leq \gamma \norm{\nabla u(x)}_{H^0(V)}^2
  \label{eq:Poincare}
\end{equation}
with $\gamma>0$ depending on $V$, is fulfilled.\\
\noindent \textbf{Cauchy-Schwarz Inequality:}
\begin{equation}
  \int_V f(x)g(x) \, dV \leq \norm{f(x)}_{H^0(V)} \norm{g(x)}_{H^0(V)},
  \label{eq:Cauchy}
\end{equation}
for any $f,g\in {\cal L}_2(V)$.

\section{Weak solution of the 3D diffusion equation}
\label{app:weak}

\begin{dfn}\cite{b:Pazy,b:PISANO2017447}
A continuous function $u(x,t) \in H^1(V)$ is said to be a weak solution of the BVP \eqref{eq:diff} on $t \geq 0$, if for every $\phi(x) \in H^1(V)$ under BCs $\phi(x)=0$ $\forall$ $x \in \partial V$, the function $\int_{V} u(x,t)\phi(x) \, dV$ is absolutely continuous on $t \geq 0$ and the relation
\begin{equation}
\begin{split}
  \int_{V} u_t(x,t)\phi(x) \, dV &= - c_{hy} \int_{V} \nabla u(x,t) \left[\nabla \phi(x) \right]^T \, dV\\ 
  &\quad + \frac{1}{\beta}[\phi(x_s^1),...,\phi(x_s^{m_s})] Q_s(t) \\
  &\quad + \frac{1}{\beta}[\phi(x_c^1),...,\phi(x_c^{m_c})] Q_c(t),
\end{split}
\label{eq:weak}
\end{equation}
holds for almost all $t \geq 0$. 
\end{dfn}

The weak solution \eqref{eq:weak} is obtained by multiplying \eqref{eq:diff} by the test function $\phi(x)$ and integrating with respect to the space variable:
\begin{equation*}
\begin{split}
  \int_{V} u_t(x,t)\phi(x) \, dV &= c_{hy} \int_{V} \nabla^2 u(x,t) \phi(x) \, dV  \\
  &\quad + \frac{1}{\beta}\int_{V}\left[\mathcal{B}_s(x) Q_s(t)\right]\phi(x) \, dV\\
  &\quad + \frac{1}{\beta}\int_{V}\left[\mathcal{B}_c(x) Q_c(t) \right]\phi(x) \, dV.
\end{split}
\end{equation*}
Using integration by parts and the definition of the Dirac's distribution, the latter expression can be rewritten as
\begin{equation*}
\begin{split}
  \int_{V} u_t(x,t)\phi(x) \, dV &= c_{hy} \int_{V} \nabla \cdot \left[\nabla u(x,t) \phi(x) \right] \, dV  \\
  &\quad - c_{hy} \int_{V} \nabla u(x,t) \left[\nabla \phi(x) \right]^T \, dV  \\ 
  &\quad + \frac{1}{\beta}[\phi(x_s^1),...,\phi(x_s^{m_s})] Q_s(t) \\
  &\quad + \frac{1}{\beta}[\phi(x_c^1),...,\phi(x_c^{m_c})] Q_c(t).
\end{split}
\end{equation*}
Finally, to retrieve expression \eqref{eq:weak} from the latter expression, the divergence theorem and the BCs were used in the first term of the RHS.

\section{Depth average of the 3D diffusion equation}
\label{app:average}

The system described by \eqref{eq:diff} is a three-dimensional system, whose solution would be difficult to plot in a simplified manner. For this purpose and without loss of generality of the theoretical results presented in this study, we chose to limit our numerical simulations to a two-dimensional boundary value problem, which was derived by depth averaging the full, three-dimensional problem given in \eqref{eq:diff} (see \cite{doi:10.1144/SP528-2022-169} for another example of depth averaging). The depth averaging was performed as follows
{\small
\begin{equation*}
\begin{split}
  \frac{1}{D_z}\int_{0}^{H} u_t(x,t) \, dx_3 &= \frac{c_{hy}}{D_z}\int_{0}^{H} \nabla^2 u(x,t) \, dx_3 \\
  &\quad + \frac{1}{\beta D_z}\int_{0}^{H}\left[\mathcal{B}_s(x) Q_s(t) \right] \, dx_3\\
  &\quad + \frac{1}{\beta D_z}\int_{0}^{H}\left[\mathcal{B}_c(x) Q_c(t) \right] \, dx_3\\
  &= \frac{c_{hy}}{D_z}\int_{0}^{H} \frac{\partial^{2} u(x,t)}{\partial x_1^{2}} \, dx_3 \\
  &\quad + \frac{c_{hy}}{D_z}\int_{0}^{H} \frac{\partial^{2} u(x,t)}{\partial x_2^{2}} \, dx_3\\
  &\quad + \frac{c_{hy}}{D_z}\int_{0}^{H} \frac{\partial^{2} u(x,t)}{\partial x_3^{2}} \, dx_3\\
  &\quad + \frac{1}{\beta D_z}\left[\mathcal{\bar{B}}_s(\bar{x}) Q_s(t)+ \mathcal{\bar{B}}_c(\bar{x}) Q_c(t) \right],
\end{split}
\end{equation*}}
where $H$ is the height of the reservoir, the new space variable $\bar{x} \in \Re^2$, $\bar{x}=[x_1,x_2]^T$. $\mathcal{\bar{B}}_s(\bar{x}) = [\delta(\bar{x}-\bar{x}_s^1),...,\delta(\bar{x}-\bar{x}_s^{m_s})]$ and $\mathcal{\bar{B}}_c(\bar{x}) = [\delta(\bar{x}-\bar{x}_s^1),...,\delta(\bar{x}-\bar{x}_c^{m_c})]$. We note that $\int_{0}^{H} \frac{\partial^{2} u(x,t)}{\partial x_3^{2}} \, dx_3=0$ due to the BC, $u(x,t) = 0 \quad \forall \quad x \in S$. Defining the depth average pressure as $\bar{u}(\bar{x},t)=\frac{1}{D_z}\int_{0}^{H} u(x,t) \, dx_3$, the last expression becomes
\begin{equation}
\begin{split}
  \bar{u}_{t}(\bar{x},t) &= c_{hy} \nabla^2 \bar{u}(\bar{x},t)+ \frac{1}{\beta D_z}\left[\mathcal{\bar{B}}_s(\bar{x}) Q_s(t)+ \mathcal{\bar{B}}_c(\bar{x}) Q_c(t) \right],\\
  \bar{u}(\bar{x},t) &= 0 \quad \forall \quad \bar{x} \in \partial S.
\end{split}
\label{eq:diff2D}
\end{equation}

Note how the systems \eqref{eq:diff} and \eqref{eq:diff2D} obtain finally the same form, allowing the theoretical developments of section \ref{app:Control} to be applied without any change. This 2D diffusion equation is numerically solved in Sections \ref{sec:motivation} and \ref{sec:sim} using the spectral decomposition presented in the following appendix.

\section{Spectral Decomposition of the 2D diffusion equation}
\label{app:spectral}

We decompose the function $\bar{u}_{t}(\bar{x},t)$ of the BVP \eqref{eq:diff2D} according to
\begin{equation}
    \bar{u}_{t}(\bar{x},t) = \sum_{\substack{n=1\\m=1}}^{\infty} z_{nm}(t) \phi_{nm}(\bar{x}),
    \label{eq:approx}
\end{equation}
where $z_{nm}(t)=\left \langle \bar{u}_{t}(\bar{x},t), \phi_{nm}(\bar{x}) \right \rangle$ is the $nm$-th Fourier coefficient of $\bar{u}_{t}(\bar{x},t)$ and $\phi_{nm}(\bar{x})$ is the $nm$-th orthonormal eigenfunction satisfying the BC. The expression $\left \langle \cdot, \cdot \right \rangle$ denotes the inner product, \textit{i.e.}, $\left \langle f(\cdot), g(\cdot) \right \rangle = \int_S f(\cdot)g(\cdot) \, dS$. For the case of the BVP \eqref{eq:diff2D}, the eigenfunction, $\phi_{nm}(\bar{x})$, and the corresponding eigenvalues $\lambda_{nm}$ are
\begin{equation}
\begin{split}
    \phi_{nm}(\bar{x}) &= 2 \sin\left( \frac{n \pi x_1}{D}\right) \sin\left( \frac{m \pi x_2}{D}\right), \\
    \lambda_{nm} &= 2 \frac{\pi^2}{D^2} \left(n^2+m^2 \right).
\end{split}
\label{eq:eigen}
\end{equation}

In order to simplify the notation, we adopt the mapping $k=h(n,m)$, which leads to the more compact form
\begin{equation}
    \bar{u}_{t}(\bar{x},t) = \sum_{k=1}^{\infty} z_{k}(t) \phi_{k}(\bar{x}).
    \label{eq:approx2}
\end{equation}
Substituting expression \eqref{eq:approx2} in \eqref{eq:diff2D} results in
\begin{equation}
\begin{split}
    \dot{z}_k(t) &= -c_{hy} \lambda_k z_k(t) + \frac{1}{\beta D_z}[\phi_k(\bar{x}_s^1),...,\phi_k(\bar{x}_s^{m_s})] Q_s(t) \\
    &\quad + \frac{1}{\beta D_z}[\phi_k(\bar{x}_c^1),...,\phi_k(\bar{x}_c^{m_c})] Q_c(t),\\
    z_{k}(0) &= \left \langle \bar{u}_{t}(\bar{x},0), \phi_{k}(\bar{x}) \right \rangle, \quad \forall \quad k\in [1,\infty).
\end{split}
\label{eq:diffmodal}
\end{equation}

Systems \eqref{eq:diff2D} and \eqref{eq:approx2}--\eqref{eq:diffmodal} are equivalent when $k \rightarrow \infty$, but the significant difference is that system \eqref{eq:diffmodal} is an ODE that can be easily implemented numerically with $k$ finite. In our numerical simulations, we were limited to 160 eigenmodes, which was more than enough according to convergence analyses. These convergence analyses are standard and were omitted from the manuscript.

\section{Output Feedback Tracking Control Design}
\label{app:Control}

The control design will be performed under the following assumptions for system \eqref{eq:diff}--\eqref{eq:SR}:
\begin{assumption}\label{A1}
The diffusion system \eqref{eq:diff} fulfils   
\begin{equation}
    \norm{\nabla^2 u_t(x,t)}_{H^0(V)} \leq L_u,
    \label{eq:ut}
\end{equation}
with known constant $L_u \geq 0$.
\end{assumption}
\begin{assumption}\label{A2}
The fixed, not controlled flux input, $Q_s(t)$, in system \eqref{eq:diff} fulfils
\begin{equation}
  \norm{\dot{Q}_s(t)} \leq L_Q,
  \label{eq:Qs}
\end{equation}
with known constant $L_Q \geq 0$.
\end{assumption}
\begin{assumption}\label{A3}
The reference to be followed, $r(t)$, is designed to fulfil
\begin{equation}
  \norm{\dot{r}(t)} \leq L_{\dot{r}}, \quad \norm{\ddot{r}(t)} \leq L_{\ddot{r}},
  \label{eq:ref}
\end{equation}
with known constants $L_{\dot{r}} \geq 0$, $L_{\ddot{r}} \geq 0$.
\end{assumption}
\begin{assumption}\label{A4}
All the parameters of the system \eqref{eq:diff}--\eqref{eq:SR} are uncertain and only nominal values are known (\textit{e.g.}, a nominal value, $f_0$, is known for the parameter $f$).
\end{assumption}

\begin{rem}
Assumption \ref{A1} is feasible due to energy conservation on the realistic system \eqref{eq:diff}. Furthermore, assumptions \ref{A2}-\ref{A4} are easily met in control applications.
\end{rem}

The first step on the design, will be to obtain the error dynamics of \eqref{eq:error} as
\begin{equation*}
\begin{split}
  \dot{y}_{e_i} &= \frac{f}{t_a \dot{\tau}_0 V_i}\int_{V_i} u_t(x,t) \, dV - \frac{1}{t_a}(e^{y_{e_i}+r_i}-1) - \dot{r}_i,
\end{split}
\end{equation*}
for $i \in [1,m_c]$. Using the 3D diffusion equation \eqref{eq:diff} and the divergence theorem, the error dynamics becomes
\begin{equation*}
\begin{split}
  \dot{y}_{e_i} &= \frac{c_{hy} f}{t_a \dot{\tau}_0 V_i}\int_{V_i} \nabla^2 u(x,t) \, dV\\
  &\quad +\frac{f}{t_a \dot{\tau}_0 \beta V_i}\int_{V_i} \left[\mathcal{B}_s(x) Q_s(t)+ \mathcal{B}_c(x) Q_c(t) \right] \, dV \\
  &\quad - \frac{1}{t_a}(e^{y_{e_i}+r_i}-1)-\dot{r}_i\\
  &= \frac{c_{hy} f}{t_a \dot{\tau}_0 V_i}\int_{V_i} \nabla^2 u(x,t) \, dV\\
  &\quad +\frac{f}{t_a \dot{\tau}_0 \beta V_i}\sum_{j=1}^{m_s} \int_{V_i} \delta(x-x_s^j) Q_{s_j}(t) dV\\
  &\quad +\frac{f}{t_a \dot{\tau}_0 \beta V_i}\sum_{j=1}^{m_c} \int_{V_i} \delta(x-x_c^j) Q_{c_j}(t) dV \\
  &\quad - \frac{1}{t_a}(e^{y_{e_i}+r_i}-1)-\dot{r}_i,\\
\end{split}
\end{equation*}
for $i \in [1,m_c]$. 

The error dynamics can be represented in matrix form as follows
\begin{equation}
  \dot{y}_e = \Psi(t) + B_s Q_s(t) + B_c Q_c(t)- \Phi(t)-\dot{r}(t),
  \label{eq:errordyn2}
\end{equation}
where $B_s=[b_{ij}^s] \in \Re^{m_c \times m_s}$, $B_c=[b_{ij}^c] \in \Re^{m_c \times m_c}$, $\Psi(t) \in \Re^{m_c}$ and $\Phi(t) \in \Re^{m_c}$ are defined as 
\begin{equation}
\begin{split}
  b_{ij}^s &= \left\{ \begin{array}{c}
  \frac{f}{t_a \dot{\tau}_0 \beta V_i} \quad \textup{if} \quad x_s^j \in V_i \\ 
  \hspace{20pt} 0 \hspace{30pt} \textup{if} \quad x_s^j \notin V_i
  \end{array} \right. ,
  \begin{array}{c}
  i \in [1,m_c] \\ 
  j \in [1,m_s]
  \end{array} , \\
  b_{ij}^c &= \left\{ \begin{array}{c}
  \frac{f}{t_a \dot{\tau}_0 \beta V_i} \quad \textup{if} \quad x_c^j \in V_i \\ 
  \hspace{20pt} 0 \hspace{30pt} \textup{if} \quad x_c^j \notin V_i
  \end{array} \right. ,
  \begin{array}{c}
  i \in [1,m_c] \\ 
  j \in [1,m_c]
  \end{array} , \\
  \Psi(t) &= \left[\begin{array}{c}
  \frac{c_{hy} f}{t_a \dot{\tau}_0 V_1}\int_{V_1} \nabla^2 u(x,t) \, dV \\ 
  \vdots \\ 
  \frac{c_{hy} f}{t_a \dot{\tau}_0 V_{m_c}}\int_{V_{m_c}} \nabla^2 u(x,t) \, dV
  \end{array}  \right], \\ 
  \Phi(t) &= \left[\begin{array}{c}
  \frac{1}{t_a}(e^{y_{e_1}+r_1}-1) \\ 
  \vdots \\ 
  \frac{1}{t_a}(e^{y_{e_{m_c}}+r_{m_c}}-1)
  \end{array}  \right],
\end{split}
  \label{eq:BPsi2}
\end{equation}
where the definition of Delta's distribution has been used.

The matrices $B_c$, $\Psi(t)$ and $\Phi(t)$ are assumed to fulfil
\begin{equation}
  B_c = \Gamma B_0, \quad \norm{\dot{\Psi}(t)} \leq L_1, \quad \norm{\dot{\Phi}(t)} \leq L_2,
  \label{eq:bounds}
\end{equation}
where $B_0 \in \Re^{m_c \times m_c}$ is a known regular matrix (consequently, $B_c$ is assumed to be regular matrix as well), $\Gamma \in \Re^{m_c \times m_c}$ is an uncertain matrix with positive diagonal entries, and $L_1 \geq 0$, $L_2 \geq 0$ are known constants. 

\begin{rem}
The assumption over the term $\Phi(t)$ in \eqref{eq:bounds} requires Assumption \ref{A3} to be fulfilled and the boundedness of the error vector derivative, $\dot{y}_e$, as $\norm{\dot{y}_e(t)} \leq L_{y_e}$, $L_{y_e} > 0$. Therefore, only local results on system \eqref{eq:errordyn2} are considered in this paper. Furthermore, the condition over the term $\Psi(t)$ requires further analysis, which will be performed in the next Lemma.
\end{rem}

\begin{lem}
\label{lemma}
The term $\Psi(t)$ in system \eqref{eq:errordyn2},\eqref{eq:BPsi2} fulfils the condition \eqref{eq:bounds} if Assumption \ref{A1} is fulfilled.
\end{lem}

\begin{pf}
Calculating the norm of the term $\Psi(t)$ defined in \eqref{eq:BPsi2} results in
\begin{equation*}
\begin{split}
    \norm{\Psi(t)} &= \frac{c_{hy} f}{t_a \dot{\tau}_0} \sqrt{\sum_{i=1}^{m_c} \frac{1}{V_i^2} \left[ \int_{V_i} \nabla^2 u(x,t) \, dV \right]^2} \\
    &\quad \leq \frac{c_{hy} f}{t_a \dot{\tau}_0} \sqrt{\sum_{i=1}^{m_c} \frac{1}{V_i^2} \int_{V_i} \left[\nabla^2 u(x,t) \right]^2 \, dV}.
\end{split}
\end{equation*}
Taking the time derivative of the last expression reads as
\begin{equation*}
\begin{split}
    \norm{\dot{\Psi}(t)} &\leq \frac{c_{hy} f}{t_a \dot{\tau}_0} \sqrt{\sum_{i=1}^{m_c} \frac{1}{V_i^2} \int_{V_i} \left[\nabla^2 u_t(x,t) \right]^2 \, dV}\\
    &\leq \frac{c_{hy} f}{t_a \dot{\tau}_0} \sqrt{\sum_{i=1}^{m_c} \frac{1}{V_i^2}}\norm{\nabla^2 u_t(x,t)}_{H^0(V)}
\end{split}
\end{equation*}
which is bounded as in \eqref{eq:bounds} if the assumption \eqref{eq:ut} is fulfilled.
\hfill $\blacksquare$
\end{pf}

The closed-loop system of \eqref{eq:errordyn2} with control \eqref{eq:Qsr} reads as
\begin{equation}
\begin{split}
  \dot{y}_e &= \Gamma \left(-K_1 \Sabs{y_e}^{\frac{1}{1-l}} + x_I \right), \\
  \dot{x}_I &= -K_2 \Sabs{y_e}^{\frac{1+l}{1-l}} + \dot{\Delta}(t),
\end{split}
\label{eq:closed2}
\end{equation}
where 
\begin{equation}
\begin{split}
  x_I &= \nu + \Delta(t), \\
  \Delta(t) &= \Gamma^{-1}\Psi(t) + \Gamma^{-1}B_s Q_s(t) -\Gamma^{-1} \Phi(t) + \left(I-\Gamma^{-1} \right)\dot{r}(t), \\
  \dot{\Delta}(t) &= \Gamma^{-1}\dot{\Psi}(t) + \Gamma^{-1}B_s \dot{Q}_s(t) -\Gamma^{-1} \dot{\Phi}(t)+\left(I-\Gamma^{-1} \right)\ddot{r}(t).
\end{split}
\label{eq:pert}
\end{equation}
The system of equations \eqref{eq:closed2}--\eqref{eq:pert} has a discontinuous right-hand side when $l =-1$ due to the definition of function $\Sabs{\cdot}^\gamma$ in \ref{app:notation}. In this special case, the solutions are understood in the sense of Filippov \cite{b:filippov}. The term $\Delta(t)$ is assumed to fulfil
\begin{equation}
  \norm{\dot{\Delta}(t)} \leq L_s, 
  \label{eq:delta}
\end{equation}
with \textit{a priori} known constant $L_s \geq 0$. This is always the case due to Assumptions \ref{A1}--\ref{A3} and \eqref{eq:bounds}. 

The tracking result for the output \eqref{eq:SR}--\eqref{eq:output} is then in force.
\begin{thm}
Let system \eqref{eq:errordyn2} assumed to fulfil \eqref{eq:ref}, \eqref{eq:bounds}, and \eqref{eq:delta}, be driven by the control \eqref{eq:Qsr} with $K_1>0$, $K_2>0$. Then, the origin of the error closed-loop system \eqref{eq:closed2}--\eqref{eq:pert}, is locally:
\begin{enumerate}
  \item Finite-time stable for any $L_s\geq0$, if $l=-1$.
  \item Finite-time stable for $L_s=0$, if $l=(-1,0)$.
  \item Exponentially stable for $L_s=0$, if $l=0$.
  \item Exponentially ISS w.r.t. $\Delta(t)$ for $L_s>0$, if $l=(-1,0]$.
\end{enumerate}
\label{th:2}
\end{thm}

\begin{pf}
Following \cite{b:9901971,b:Mathey-Moreno-2022}, the trajectories $(y_e,x_I)$ of system \eqref{eq:closed2}--\eqref{eq:pert} are ensured to reach the origin if $K_1$ is a matrix with positive diagonal elements and $K_2$ is a positive quasi-definite matrix. 
\hfill $\blacksquare$
\end{pf}

\begin{rem}
As a consequence of the stability of the closed-loop system trajectories, $(y_e,x_I)$, and due to the definition of the perturbation term $\Delta(t)$ in \eqref{eq:pert}, the integral term, $\nu$, of the control \eqref{eq:Qsr} is able to provide an estimate of such term, \textit{i.e.}, $\nu(t) \rightarrow -\Delta(t)$ as $t \rightarrow \infty$.
\end{rem}

\subsection{Energy demand and production constraints}
\label{app:demand}

For the case where additional number of flux restrictions to the fluid injection of the controlled injection points are considered, the vector of controlled injection points of system \eqref{eq:errordyn2} is defined as follows
\begin{equation}
  \dot{y}_e = \Psi(t) + B_s Q_s(t) + \bar{B}_c \bar{Q}_c(t) - \Phi(t)-\dot{r}(t),
  \label{eq:errordyn3}
\end{equation}
where $\bar{Q}_c(t) \in \Re^{m_c+m_r}$, $\bar{Q}_c(t)=[\bar{Q}_{c_1}(t),...,\bar{Q}_{c_{m_c+m_r}}(t)]^T$ are the new controlled fluxes and $\bar{B}_c=[\bar{b}_{ij}^c] \in \Re^{m_c \times (m_c+m_r)}$ is the new input matrix defined as 
\begin{equation}
\begin{split}
  \bar{b}_{ij}^c &= \left\{ \begin{array}{c}
  \frac{f}{t_a \dot{\tau}_0 \beta V_i} \quad \textup{if} \quad x_c^j \in V_i \\ 
  \hspace{12pt} 0 \hspace{26pt} \textup{if} \quad x_c^j \notin V_i
  \end{array} \right. ,
  \begin{array}{c}
  i \in [1,m_c] \\ 
  j \in [1,m_c+m_r]
  \end{array} , 
\end{split}
  \label{eq:BPsi3}
\end{equation}
where $x_c^j$, $j\in [1,m_c+m_r]$, are the injection points over the total region $V$. If we replace \eqref{eq:Qcr} in \eqref{eq:errordyn3}, the link between the new input matrix, $\bar{B}_c$, and the original input matrix, $B_c$, defined in \eqref{eq:BPsi2} is stated as $\bar{B}_c \overline{W} =B_c$.

\end{document}